\documentclass[journal=jacsat,manuscript=article]{achemso}

\usepackage[T1]{fontenc} 

\usepackage{graphicx}
\usepackage{color}
\usepackage{multirow}
\usepackage{physics}
\usepackage{nameref}
\usepackage{amsmath}

\usepackage{hyperref}

\author{Giorgio Visentin}
\affiliation{Department of Physics and Astronomy, Louisiana State University, Baton Rouge, Louisiana 70803, USA}
\email{gvisentin@lsu.edu}

\author{Fran\c{c}ois Mauger}
\affiliation{Department of Physics and Astronomy, Louisiana State University, Baton Rouge, Louisiana 70803, USA}
\email{fmauger@lsu.edu}

\title{Configuration-interaction calculations with density-functional theory molecular orbitals for modeling valence- and core-excited states in molecules}


\begin{document}


\begin{abstract}
    We investigate configuration-interaction (CI) calculations on a basis of molecular orbitals generated by preliminary density-functional theory (DFT) calculations. We use this CI/DFT framework to improve the modeling of core-excited states by exploiting the flexibility and account for electron correlation of DFT orbitals compared to the canonical Hartree-Fock analogs. We assess the performance of our approach on the valence- and core-excited electronic states of three molecules with increasing levels of electron-correlation complexity: the singly bonded CH$_4$, doubly bonded CO$_2$, and triply bonded N$_2$. 
    For molecules with strong electron correlation effects, such as CO$_2$ and N$_2$, the inclusion of double excitations is important to model the core-hole excited states with reasonable accuracy. CI/DFT outperforms standard single-reference CI on Hartree-Fock molecular orbitals and competes with multi-reference CI calculations with multi-configuration self-consistent field orbitals in the modeling of molecules with strong electron-correlation effects, but weak multi-reference nature of their wavefunction such as CO$_2$. In contrast, the choice of the molecular-orbital basis is irrelevant when modeling systems with negligible electron-correlation effects like CH$_4$ or important multi-reference nature of their wavefunction like N$_2$.
\end{abstract}



\section{Introduction} \label{Introduction}

The modeling of the electronic structure of molecules, including the manifold of excited states, is of great importance in quantum chemistry and molecular physics. Excited states
define the optical and reaction properties of atoms and molecules. As such, they
are important players in many field of theoretical and experimental physical chemistry such as time-resolved spectroscopy~\cite{Gonzalez:2012} and ultrafast processes~\cite{Pohl:2017, Klinker:2018, Giri:2023, Travnikova:2025} like charge migration~\cite{Unger:2017,Schlegel:2023, Hamer:2024, Folorunso:2023, Hamer:2025, Mauger:2025} and photofragmentation~\cite{Travnikova:2025, Ganguly:2023, Schewe:2022} to name a few. 
The development of fast x-ray sources, both in table-top setup~\cite{Barletta:2010} and at free-electron laser facilities~\cite{Patrick:2018, Grubel:2008, Barletta:2010}, has drawn increasing attention to core excited states and their strong element specificity~\cite{Patrick:2018}.
However, modeling of excited states in polyatomic molecular systems is challenging to accurately capture intricate electron correlation effects~\cite{Gonzalez:2012, Patrick:2018}. 
The difficulty is exacerbated for core excited levels, where the core vacancy typically leads to significant reshaping of the rest of the wave function as compared to its ground and low-lying excited states~\cite{Patrick:2018}, and requires specific computational treatments~\cite{Patrick:2018, Besley:2009, Imamura:2007, Bagus:1971, Gilbert:2008}. 

In quantum chemistry, electron correlation specifically refers to the portion of the Coulomb repulsion between electrons that is neglected in the Hartree-Fock (HF) framework.
It is commonly separated between \textit{dynamic correlation}, which refers to the instantaneous electron-electron repulsion, and \textit{static correlation} that arises when the electronic state cannot be described with a single Slater determinant \cite{Vogiatzis:2015}.
Several post-HF \textit{ab} \textit{initio} methods have been devised and model electronic structures in terms of either the electron density or the wavefunction of the system. The main methods for the former class are density functional theory (DFT) and its time-dependent extension (TDDFT), which account for electron correlation by means of specific functionals choice~\cite{Vogiatzis:2015, Patrick:2018}. One of the main advantage of (TD)DFT is its scalability~\cite{Grimme:1996} to large systems that are out of reach from most wavefunction-based \textit{ab initio} methods~\cite{Gonzalez:2012}. 
On the other hand, wavefunction-based methods typically model electron correlation as excitations from one or more references to excitation configurations, derived from a basis of molecular orbitals~\cite{Vogiatzis:2015}. Configuration-interaction (CI) is one of the most widespread wavefunction-based levels of theory due to its flexible modeling of electron correlation~\cite{Durden:2024, Saito:2021, Schlegel:2023}. In its standard implementations, CI uses a basis of HF molecular orbitals. The main drawbacks of this basis consist in the slow convergence of the related CI computations~\cite{Duch:1991}, which have spurred theorists to develop optimized-orbital frameworks such as multi-configurational self-consistent field (MCSCF) or complete active-space self-consistent field (CASSCF) methods \cite{Patrick:2018}. These bases are optimized for the modeling of static correlation, but their generation are computationally expensive~\cite{Duch:1991} and require physico-chemical intuition to reasonably model correlation effects \cite{Gonzalez:2012}.

Starting from the 1990's, theorists have been aiming at combining electron-density- and wavefunction-based levels of theory, by matching CI restricted to single excitations (CIS) with DFT molecular orbitals. The first noticeable attempt came from Grimme \cite{Grimme:1996} in 1996, where the CI matrix features HF elements corrected with DFT and empirical scaling parameters \cite{Grimme:1996}. A further step in this direction was later made by Hermann \textit{et} \textit{al.} \cite{Hermann:2017}, with the hybrid TDDFT/CIS. Unlike Grimme's model, this method does not imply empirical parameters, and generates molecular orbitals and CI expansion coefficients after a preliminary TDDFT routine. The CI Hamiltonian is then replaced with the TDDFT one, but does not explicitly retrieve excitations higher than singles, except indirectly \cite{Hermann:2017}.
Alternatively, in this work we investigate an approach that matches CI and DFT. Our CI/DFT starts from the standard CI assumption that the wavefunction can be expanded into a set of configurations but writes them in a basis of molecular orbitals obtained from an independent DFT computation. Furthermore, CI/DFT is based on the full Schr\"{o}dinger Hamiltonian rather than the (TD)DFT ones and it explicitly models the configuration basis, thus allowing for the direct retrieval of the excitation levels higher than the singles.

Our work is organized as follows: 
First, in the Model section, we outline the theoretical model underlying the CI/DFT method and tests the convergence of the ground state in the few-electron LiH molecule with the size of the configuration basis.
Next, in the Results and discussion section, we assess the performance of CI/DFT in three closed-shell molecules with increasing electron-correlation complexity: CH$_4$, with only single covalent bonds; CO$_2$, with two double covalent bonds; and N$_2$, with one triple covalent bond. For each target, we benchmark the CI/DFT vertical excitation energies associated with valence- and core-hole excited states against experimental and high-level {\it ab-initio} calculation available in the literature.
Finally, we summarize our results in the Conclusions section.

\section{Model} \label{sec:Model}

In this section, we outline the theoretical and computational background of our CI/DFT approach.
First, we summarize the main equations underpinning the approach and how we build the CI-Hamiltonian matrix for multi-active electron systems. 
Next, we discuss the implementation of CI/DFT using the Psi4 quantum chemistry package~\cite{Psi4}.
Finally, we describe a pristine test of CI/DFT on the ground-state energy convergence of the four-electron LiH molecule. 

\subsection{Theoretical background} \label{sec:Theorback}

Similar to conventional CI models, the CI/DFT method aims to solve the non-relativistic time-independent Schr\"{o}dinger equation
\begin{equation}\label{eq:Schroedinger}
    \hat{H} \ket{\Psi} = E \ket{\Psi},
\end{equation}
where $\hat{H}$ is the multi-electron Hamiltonian operator, and~$E$ and~$\ket{\Psi}$ are a pair of eigen-value and eigen-vector of the Hamiltonian, expressed in a basis of antisymmetrized products of one-electron wave functions independently obtained from a DFT calculation.
In fact, CI/DFT differs from standard CI on HF molecular orbitals for the choice of the molecular-orbital basis.

For a N-electron molecular system, the clamped-nuclei Hamiltonian $\hat{H}$ in equation~\eqref{eq:Schroedinger} reads
\begin{equation}\label{eq:H}
    \hat{H}({\bf x}_1, {\bf x}_2, ..., {\bf x}_N ) =
        -\sum_{k=1}^{N} \frac{\Delta_k}{2}
        + \sum_{k=1}^{N} V_{\text{ne}}({\bf r}_k)
        + \sum_{1 \leq k < l \leq N} V_{\text{ee}}({\bf r}_k - {\bf r}_l),
\end{equation}
where ${\bf x}_k=({\bf r}_k,\omega_k)$ are the electronic coordinates including the spatial ${\bf r}_k$ and spin $\omega_k$ information, and $\Delta_k$ is the Laplacian operator with respect to the electronic coordinate~\textit{k}. 
The first two sums in the equation correspond to the one-electron operator, respectively including the kinetic and electron-nucleus interactions via the attractive potential $V_{\text{ne}}$.
The final sum corresponds to the two-electron operator, grouping all the electron-electron interactions via the Coulomb potential~$V_{\text{ee}}({\bf r})=1/|{\bf r}|$. 

We expand the wavefunction $\ket{\Psi}$ of equation~\eqref{eq:Schroedinger} as a linear combination of Slater Determinants~$\ket{\psi_{\bf k}}$, called the \emph{configurations} of the system~\cite{Szalay:2011, Duch:1991}. 
Specifically, each configuration is obtained by selecting a set of~$N$ unique orbitals out of an orthonormal set of~$K>N$ one-electron spin orbitals~$\{\chi_k\}_k$, for us obtained from an independent DFT calculation, and forming the antisymmetrized product
\begin{equation} \label{eq:model:Slater_determinant}
    \ket{\psi_{\bf k}} = |\chi_{k_1} \chi_{k_2} \ldots \chi_{k_N}\rangle = 
        \frac{1}{\sqrt{N!}}\sum_{\sigma\in S_N}{\text{sgn}(\sigma) \prod_{n=1}^{N}{\chi_{\sigma(l)}({\bf x}_{k_n})}}.
\end{equation}
Here~$S_N$ is the group of permutations in~$[1,N]$ and~$\text{sgn}(\sigma)$ is the signature of the permutation~$\sigma$.
The accuracy of CI/DFT models is thus determined by (i) the choice of the DFT functional and, for basis-set calculations, the basis of atomic orbitals used to calculate the spin orbitals, (ii) the number of those spin orbitals allowed in the definition of the configuration states, and (iii) the subset of configurations actually included in the CI/DFT expansion. 
Throughout this Paper, we label configurations by the $N$-element index vector ${\bf k}=\{k_{l_1},k_{l_2},\ldots, k_{l_N}\}$ of the spin orbitals $\chi_{l_1}$, $\chi_{l_2}$, \ldots $\chi_{l_N}$ they employ. Once the configuration basis is selected, we rewrite the eigen-problem of equation~\eqref{eq:Schroedinger} in the span of the basis by forming the CI/DFT Hamiltonian matrix with elements 
\begin{equation} \label{eq:CI-DFT_Hamiltonian}
    H_{{\bf k},{\bf k}^\prime} = 
        \langle \psi_{\mathbf{k}} | {\hat{H}} | \psi_{\mathbf{k}^\prime} \rangle,
        \quad
        \forall {\bf k}, {\bf k}^\prime,
\end{equation}
and solve for the spectrum of the matrix $H$.
To streamline notations in the expressions for these matrix elements, we introduce the core-integral between two spin orbitals labeled by their indexes $k$ and $k^\prime$
\begin{equation} \label{eq:core_integral}
    \langle k | \hat{h} | k^\prime \rangle = 
        \int \chi_k({\bf x})^* \hat{h}({\bf r}) \chi_{k^\prime}({\bf x})\, \text{d}{\bf x}
        \quad
        \text{with}
        \quad
        \hat{h}({\bf r}) = -\frac{\Delta}{2}+V_{\text{ne}}({\bf r}),
\end{equation}
and the two-electron integral between four spin orbitals
\begin{equation} \label{eq:two-electron_integral}
    \langle k k^\prime | l l^\prime \rangle = 
        \iint \chi_k({\bf x})^* \chi_{k^\prime}({\bf x^\prime})^*
            V_{\text{ee}}({\bf r} - {\bf r}^\prime) 
            \chi_l({\bf x}) \chi_{l^\prime}({\bf x^\prime})\,
        \text{d}{\bf x} \, \text{d}{\bf x}^\prime.
\end{equation}
The matrix elements of equation~\eqref{eq:CI-DFT_Hamiltonian} are evaluated by means of the Slater-Condon rules~\cite{SzaboBOOK, Slater:1929, Condon:1930}, which discriminate four different cases:
the (i) diagonal elements of the CI/DFT matrix correspond to the self pairing,
\begin{equation} \label{eq:Model:CI_matrix_diagonal}
    H_{{\bf k}, {\bf k}} = 
    \sum_{n=1}^{N}{ \langle k_{l_n} | \hat{h} | k_{l_n} \rangle} 
    + \frac{1}{2} \sum_{n, n^\prime=1}^{N}{ 
        \langle k_{l_n} k_{l_{n^\prime}} | k_{l_n} k_{l_{n^\prime}} \rangle 
        - \langle k_{l_n} k_{l_{n^\prime}} | k_{l_{n^\prime}} k_{l_n} \rangle 
    };
\end{equation}
(ii) off-diagonal elements with configuration pairings that differ by a single spin orbital,
\begin{equation} \label{eq:Model:CI_matrix_single}
     H_{{\bf k},{\bf k}_a^r} = 
        \langle k_a | \hat{h} | k_r \rangle + \sum_{n=1}^{N}{\langle k_a k_{l_n} | k_r k_{l_n} \rangle - \langle k_a k_{l_n} | k_{l_n} k_r \rangle} ,
\end{equation}
where ${\bf k}_a^r$ indicates that the spin orbital with index~$a$ is replaced with that of index~$r$ in~${\bf k}$ while the other ones remain unchanged\footnote{In general, writing ${\bf k}^\prime={\bf k}_a^r$ in the matrix element of equation~\eqref{eq:CI-DFT_Hamiltonian} involves some permutation of the spin-orbital indexes. This is done using the antisymmetric property of configuration states of equation~\eqref{eq:model:Slater_determinant}.}; 
(iii) off-diagonal elements with two spin orbital differences,
\begin{equation} \label{eq:Model:CI_matrix_double}
     H_{{\bf k},{{\bf k}_{ab}^{rs}}} = 
        \langle k_a k_b | k_r k_s \rangle - \langle k_a k_b | k_s k_r \rangle,
\end{equation}
where orbitals $a$ and $b$ are replaced with $r$ and $s$, respectively;
And~(iv) all the other pairings, involving pairs of configurations with more than two spin orbital differences that all vanish. 

We follow conventional CI nomenclature to define the choice of configuration bases in our CI/DFT calculations. 
If the wavefunction $\Psi$ is written in the basis of all possible configurations, the diagonalization of the
matrix $H$ is equivalent to the exact solution of the time-independent Schr\"{o}dinger equation (\ref{eq:Schroedinger}) \cite{SzaboBOOK, ForesmanBOOK} and the method is called~\emph{full-CI}~\cite{Szalay:2011, Duch:1991}. However, full-CI calculations are computationally unfeasible for most atomic and molecular systems and approximate CI wavefunctions are often used as a reasonable compromise between accuracy and computational time. The idea behind such approximations is that many pairings provide small contributions to the modeling of electron correlation~\cite{Vogiatzis:2015} and therefore can be removed by means of several viable strategies. 
One of these strategies truncates the configuration basis to a specific excitation level by defining so-called \emph{reference} and \emph{excitation spaces}. The excitation space consists of those configurations that are obtained upon replacement of one or more occupied spin orbitals in a reference with unoccupied (\textit{virtual}) orbitals. The resulting configurations are called \emph{excitations} and are hierarchically distinguished in terms of the maximum number of replacements, as \emph{single} excitations or \textit{singles} (CIS, with one replacement), \emph{double} excitations or \emph{doubles} (CISD, two replacements), \emph{triple} excitations or \emph{triples} (CISDT, three replacements) and so forth~\cite{Szalay:2011, SzaboBOOK}. When the reference space is limited to a single configuration the calculation is said to be \emph{single reference} (SR-CI or simply CI). Otherwise it is called \emph{multi reference} (MRCI)~\cite{Szalay:2011}.
A second strategy consists of restricting the set of orbitals used in the calculations to those that generate the most important configurations~\cite{Vogiatzis:2015, Casanova:2022}. \emph{Complete active space} (CAS) models then form all possible configurations with the selected orbitals. The active space (AS) can be further partitioned into subspaces in which selective excitation levels are allowed~\cite{Casanova:2022, Duch:1991} and are called \textit{restricted active spaces} (RAS) within the AS. 
In this Paper, we carry out single-reference, multi-reference, as well as RAS-type CI/DFT calculations.

\subsection{Implementation} \label{sec:Implementation}

All the numerical results reported in this Paper rely on the \texttt{Psi4} quantum-chemistry package~\cite{Psi4}. 
Specifically, we use the built-in self-consistent-field routines to determine the spin-restricted HF and DFT molecular orbitals that define the spin-orbital basis.
We also rely on \texttt{Psi4}'s routines to obtain the core matrix and electron-repulsion four-tensor, respectively defined by equations~\eqref{eq:core_integral} and~\eqref{eq:two-electron_integral} over all possible spin-orbital indexes used in the configuration basis.
From these, we build the CI/DFT Hamiltonian matrix of equations~\eqref{eq:CI-DFT_Hamiltonian} using the Slater-Condon rules of equations~(\ref{eq:Model:CI_matrix_diagonal}-\ref{eq:Model:CI_matrix_double}). Note that, because of the restricted ground-state calculation, up- and down-spin orbitals share the same spatial-orbital components, which we capitalize on when building the Hamiltonian matrix.
Finally, we obtain the ground- and excited-state energies together with the wave functions by diagonalizing the thus-obtained Hamiltonian matrix.
Our CI/DFT code \texttt{psiCI} is available on the GitHub repository~\cite{psiCI}, which also includes a documentation and a handful of examples for how to set calculations.
We have checked the accuracy of our CI/DFT implementation by performing CI calculations with HF molecular orbitals, for which closed formula expressions for the CI matrix elements are known~\cite{SzaboBOOK}.
We note that even at the CIS level and with explicitly calculating matrix elements associated with the pairing between the reference and single-excitation states, all four Slater-Condon rules are used to build the Hamiltonian matrix and thus provide a complete test for \texttt{psiCI}.
We have computed the DFT molecular orbitals with the generalized-gradient-approximation (GGA) BLYP functional~\cite{Becke:1988, Lee:1988} alongside the two PBE0~\cite{PBE0} and B3LYP~\cite{B3LYP_1, B3LYP_2} hybrid functionals, which are widely employed in the literature. We have generally found that CI/DFT-PBE0 and CI/DFT-B3LYP give very similar results and yield curves that are between CI/HF and CI/DFT-BLYP. We speculate that feature to stem from the fraction of HF included in the two hybrid functionals. To enhance the readability of figures, throughout the Paper we report the results obtained using BLYP molecular orbitals and provide extended results including the PBE0 and B3LYP functionals in the supporting information.

\subsection{Test case: LiH molecule ground state} \label{sec:Testcase}

\begin{figure}[htb]
    \centering
    \includegraphics[width=0.75\linewidth]{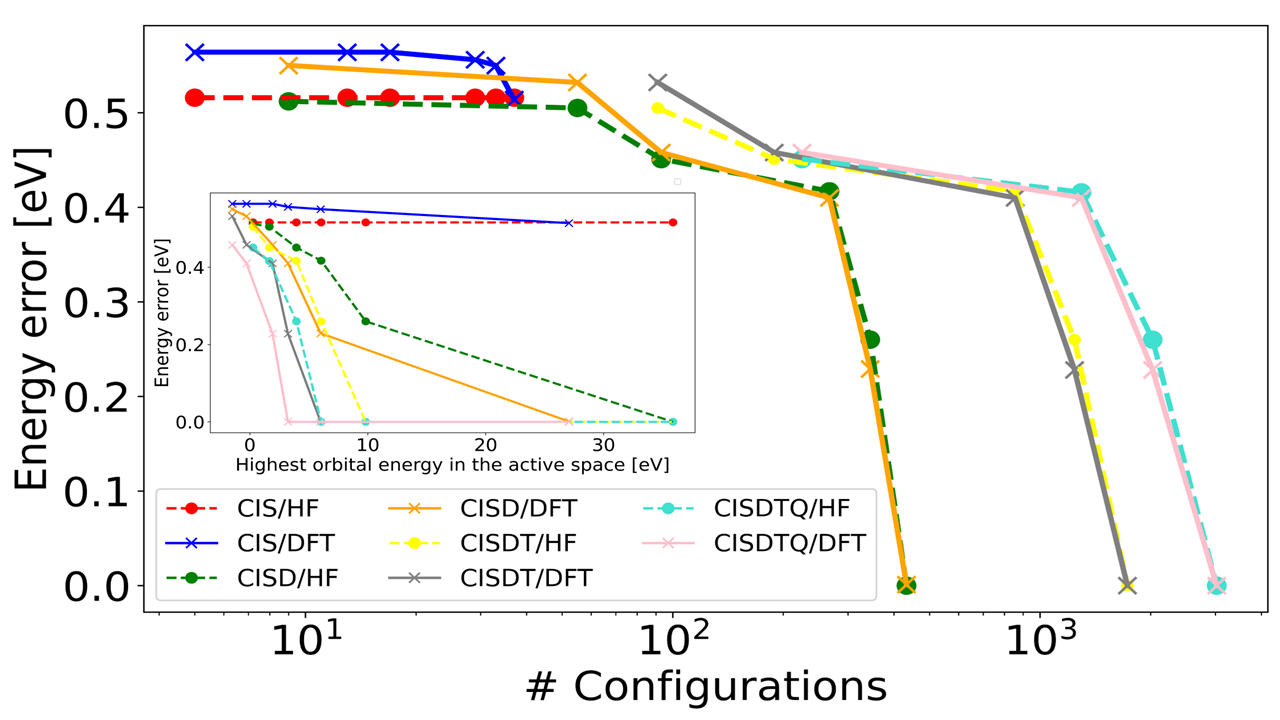}
    \caption{\label{LiH}
    Convergence pattern for the ground state of LiH in terms of the (main panel) number of configurations and (inset) energy of the highest-lying molecular orbital included in the calculation's active space -- see legend. 
    For each calculation we define the energy error as the difference between the lowest eigen value of the CI/DFT Hamiltonian matrix and that of the full-CI limit for the same Gaussian 6-31G basis.
    }
\end{figure}

We first investigate the behavior of our CI/DFT method for the convergence of the ground-state's electronic energy of LiH. This small four-electron molecule allows us to systematically study its ground-state energy convergence with respect to increasing the (i) excitation levels and (ii) size of the active space up to the full-CI limit. 
To this end, we performed computations using both HF and DFT molecular orbitals, with the small Gaussian 6-31G \cite{6-31g} basis set. 
We then used these two orbital bases to compute the ground-state energy at the CIS, CISD, CISDT and CISDTQ levels -- this latter being equivalent to a full-CI computation -- and on active spaces including virtual orbitals of incrementally higher energies. 
We plot the outcome of our calculations in Figure~\ref{LiH}, where we quantify the convergence of the ground-state energy with respect to the full-CI limit for the same Gaussian 6-31G basis. The main panel and inset respectively show the results as functions of the number of configurations and the HF/DFT orbital energy of the highest-lying virtual molecular orbital included in the AS.

At the full-CI limit, we find that both HF and DFT orbital bases produce the same results -- see the lower-right most markers in the main panel and inset of figure~\ref{LiH}. This is expected as both the HF and DFT operators are Hermitian and thus their spectra span the entire vector space formed by the atomic-orbital basis.
Instead, the choice of one set of orbitals vs the other, together with the space of configuration, should aim to optimize the accuracy of the CI(/DFT) results while keeping a reasonable computational cost.
This is in general not a trivial task as the full-CI limit is inaccessible for most molecular systems larger than LiH and one would hope to get a suitable result without having to perform expensive calculations with several orbital sets. This Paper aims to provide insight around this question, with putting a special focus on core-excited levels.
For the ground-state LiH molecule of figure~\ref{LiH}, we observe only minor difference in the error between the HF and DFT molecular bases, except from CIS calculations where HF systematically outperforms its DFT alternative. We attribute these results to the very weak correlation nature of the LiH ground-state wave function.

\section{Results and discussion} \label{sec:Results_and_discussion}

\begin{figure}[htb]
    \centering
    \includegraphics[width=0.75\linewidth]{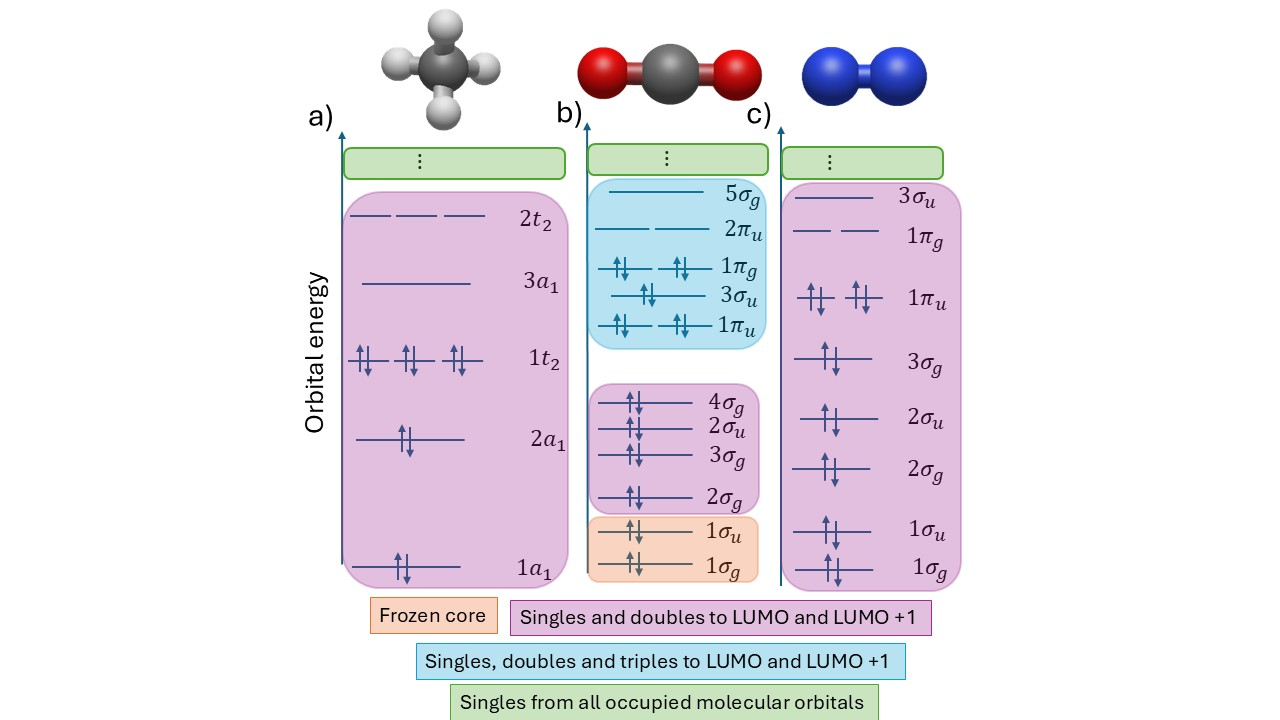}
    \caption{\label{Active_space}
    Sketches of the restricted active spaces (RAS) we use in our calculations of the electronic structures for a) CH$_4$, b) CO$_2$ and c) N$_2$. The different colored boxes labels the levels of excitations we include in the AS -- see legend. The up/down arrows specify the DFT-ground state orbital occupation.}
\end{figure}

In this section, we investigate the performance of our CI/DFT approach in the modeling of the valence- and core-excited states for three closed-shell molecules with increasingly challenging electronic structures: CH$_4$, CO$_2$, and N$_2$ -- see the sketches in figure~\ref{Active_space}. For each system, we first optimize the ground-state geometry at the MP2 level of theory and with the correlation-consistent aug-cc-pvTZ~\cite{aug-cc-pvTZ_1, aug-cc-pvTZ_2} basis set using Psi4~\cite{Psi4}.  
We then use the optimized geometries to generate a basis of HF and DFT molecular orbitals on the same atomic-orbital basis set. 
Subsequently, we run CI/HF and CI/DFT calculations with different excitation levels on the molecules' ground, valence- and core-excited states, and compute the related excitation energies $T$.  Our calculations explore the sensitivity of CI/DFT with respect to (i) the size of the AS, (ii) the molecular-orbital basis, and (iii) the excitation level within the CI expansion, alongside its accuracy with respect to state-of-the-art literature data.

\subsection{Ground, valence-, and core-excited states of CH$_4$}

We begin our analyses with the singly-bonded methane molecule CH$_4$. Its ground-state wave function belongs to the X$^1$A$_1$ point-group symmetry and we show its electronic configuration, together with the active space (AS) partitioning we use in our investigations, in  
Figure~\ref{Active_space}~a). Specifically, our minimal AS includes all the occupied molecular orbitals -- filled with up/down arrows in the figure -- plus the lowest-virtual molecular orbital (LUMO) 3a$_1$ and triply-degenerate LUMO+1 2t$_2$. We then systematically increase the size of the AS up to the maximal one, including all the molecular orbitals generated from the HF/DFT orbital basis set. 
Taking the occupied orbitals as the single reference we consider two flavors of excitations within the hierarchies of AS with (i) CIS calculations and (ii) RAS-CISD, which includes double excitations within the minimal active space and single excitations beyond.

\begin{figure}[htb]
    \centering
    \includegraphics[width=0.75\linewidth]{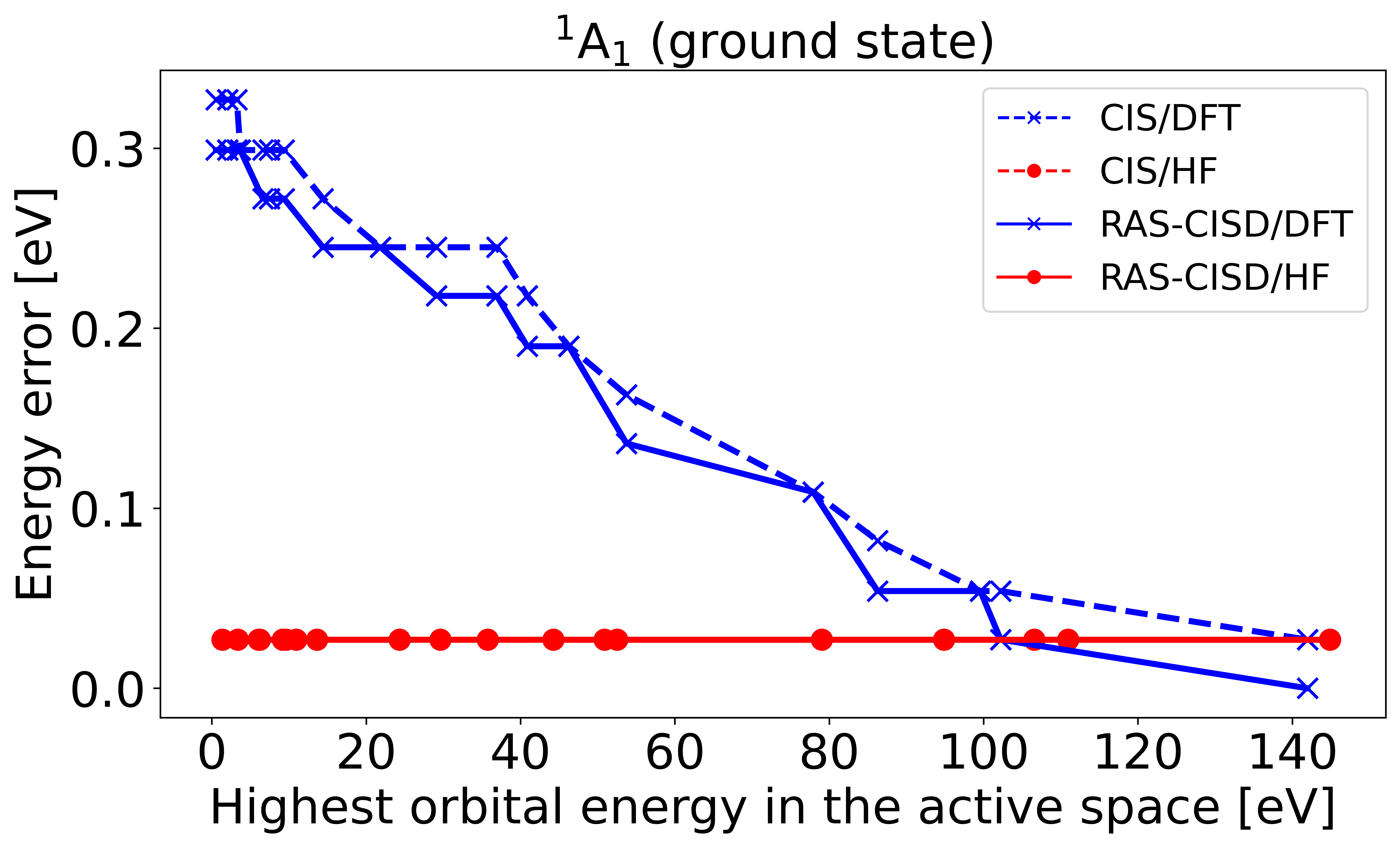}
    \caption{\label{CH4_Convergence}
    Convergence pattern for the electronic energy associated with
    the ground state of CH$_4$ as a function of the energy of the highest-lying molecular orbital included in the calculation’s active space.
    The CIS/HF and RAS-CISD/HF curves are on top of each other.
    Similar to Figure~\ref{LiH}, for each calculation we define the energy error as the difference between the lowest eigen value of the CI(/DFT) Hamiltonian matrix and the absolute lowest energy across all calculations.
    }
\end{figure}

Figure~\ref{CH4_Convergence} compares the convergence pattern for the ground state of CH$_4$ for the different HF/DFT and CIS/RAS-CISD models, as a function of the energy of the highest virtual orbital included in the AS. Using the energy-minimization property of the Schr\"{o}dinger-equation ground state, we evaluate the convergence error with respect to the minimal-energy result across all calculations.
As expected, because of Brillouin's theorem~\cite{SzaboBOOK}, CIS on a basis of HF orbitals shows no improvement compared to the HF ground state and produces a flat curve.
Similarly, we find negligible improvement with the inclusion of doubles as the RAS-CIS/HF results lie on top of the CIS/HF one.
On the other hand, for the DFT basis, we observe steady improvement from the minimal-AS result with increasing the AS, by about 0.3~eV at the full basis limit. We also notice that the inclusion of the doubles generally slightly improves the results. 
Comparing the HF and DFT curves, the latter outperforms only for the largest orbital space and at the RAS-CISD level. 
We ascribe these results to the negligible electron correlation in CH$_4$, such that the HF configuration provides an excellent description of the ground-state wave function. In turn, it explain the slower ground-state energy convergence at the CI/DFT levels compared to the CI/HF equivalents, like in the LiH case of Figure~\ref{LiH}

\begin{figure}[htb]
    \centering
    \includegraphics[width=0.75\linewidth]{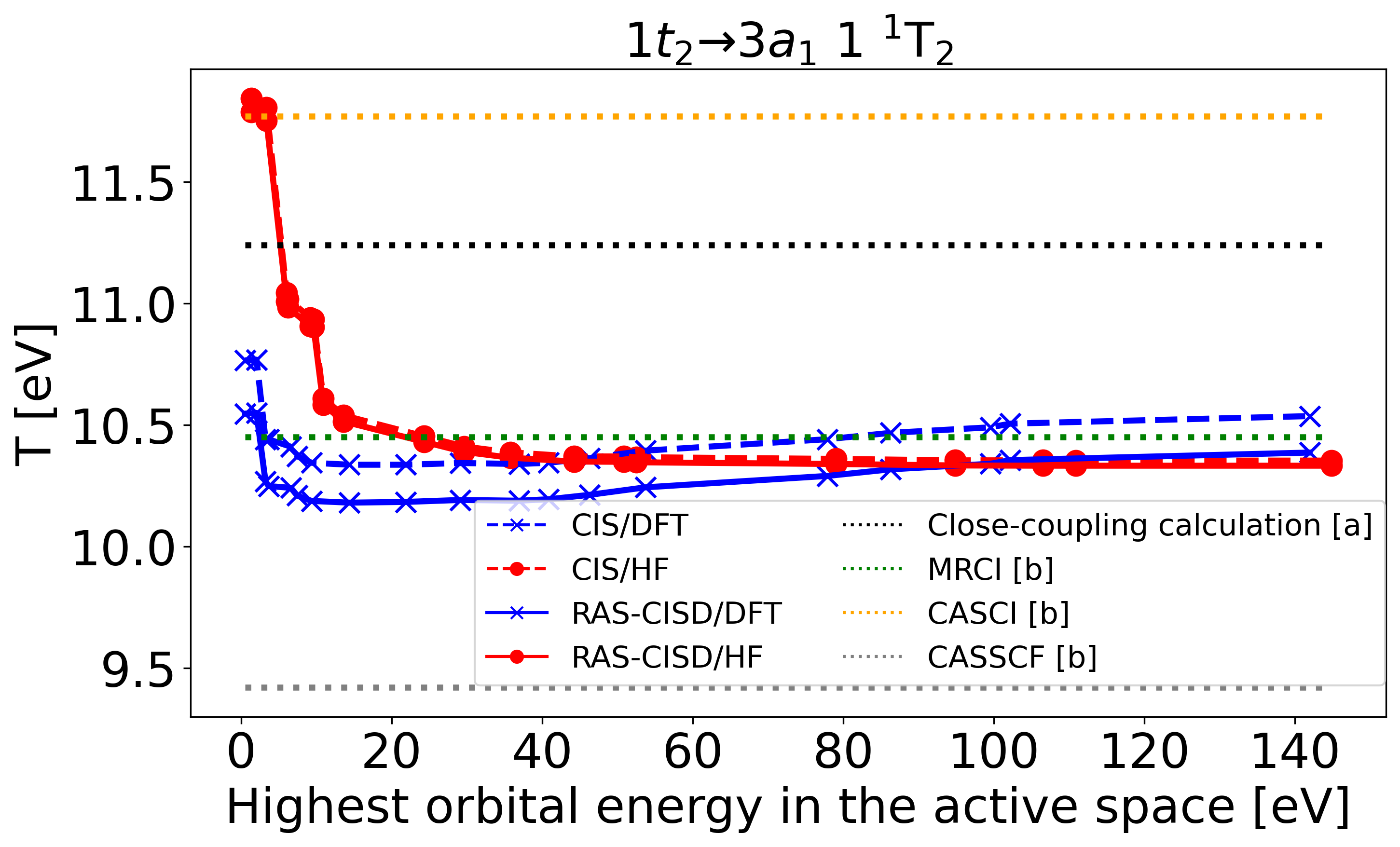}
    \caption{\label{CH4_Valence}
    Vertical excitation energy T to the first valence-excited state 1$^1$T$_2$ of CH$_4$ computed at the CIS and RAS-CISD levels on HF and DFT molecular orbital bases reported in terms of the binding energy associated with the highest-lying virtual molecular orbital included in the active space. 
    The CIS/HF and RAS-CISD/HF curves are on top of each other.
    For comparison, we also report the results of [a] complex Kohn calculations by Gil \textit{et al.}~\cite{Gil:1994} and [b]  MRCISD, CASCI and CASSCF computations by Zi\'{o}\l{}kowski \textit{et al.}~\cite{Ziolkowski:2012}.}
\end{figure}

Next, we look at the lowest-lying valence-excited state, with 1$^1T_2$ symmetry.
This state stems from the $1t_2 \to 3a_1$ electron excitation and possesses Rydberg features~\cite{Winstead:1993, Gil:1994, Ziolkowski:2012}. For the CI(/DFT) calculations, we use the same optimized geometry and HF/DFT molecular orbitals as for the molecular ground state. 
Figure~\ref{CH4_Valence} compares the vertical excitation energy X$^1$A$_1 \to 1 ^1T_2$ with theoretical results from the literature and Table~\ref{Table:CH4} summarizes the results for the largest AS.
The excited-state electronic energy is more sensitive to the AS size compared to the ground-state analog. Extension of the AS size decreases the CI/HF electronic energies by more than 1.4 eV, while the CIS and RAS-CISD/DFT counterparts are lowered by, at most, 0.7 and 0.5 eV, respectively. For all the CI approaches, the convergence plateau is attained when virtuals lying higher than 50 eV are included in the AS.
At the convergence limit, our computations are bounded from below and above by the MRCI and close-coupling references, respectively, regardless of the excitation level and molecular-orbital basis employed. In particular, the complete-active-space self-consistent-field (CASSCF)~\cite{Ziolkowski:2012} data always lie below our computations, whereas the complete-active-space CI (CASCI) from the same reference lie above them regardless of the AS size. These features show that the modeling of dynamic electron correlation in this molecule is more important than the description of static electron correlation. However, our computations are negligibly improved by the inclusion of the doubles, regardless of the molecular-orbital basis and especially at the largest AS. The CI/HF computations are particularly emblematic of this, as the CIS and RAS-CISD results overlap throughout most of the active spaces. In addition, at the largest AS the choice of the molecular orbital basis yields small differences between our two sets of RAS-CISD calculations. Like for the ground state, we ascribe this behavior to the negligible dynamic electron correlation that affect CH$_4$.

\begin{figure}[H]
    \centering
    \includegraphics[width=0.75\linewidth]{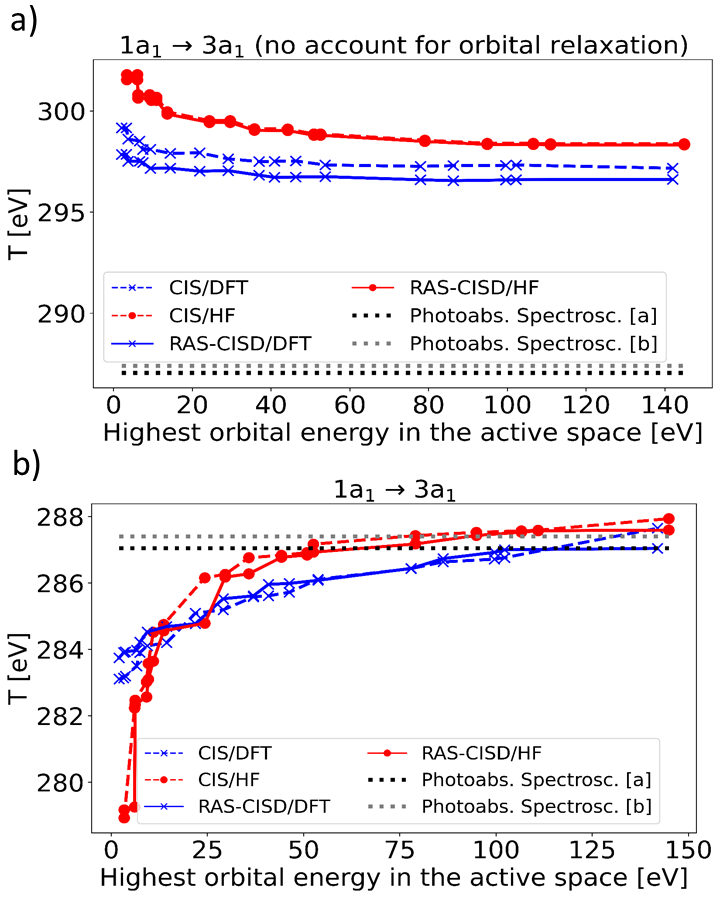}
    \caption{\label{CH4_Core}
    Convergence pattern associated with the 1$a_1$ $\to$ 3$a_1$ carbon K edge excitation energy T of CH$_4$ in terms of the active-space size for the a) non-relaxed and b) relaxed molecular-orbital bases. We benchmark our results with respect to the Photoabsorption-spectroscopy experiments by [a] Ueda \textit{et al.} \cite{Ueda:1995} and [b] Kivimaki \textit{et al.} \cite{Kivimaki:1996}}
\end{figure}

To conclude our analysis of the CH$_4$ molecule, we investigate the modeling of core-excited states. We focus on the lowest-lying excited state that stems from the $1a_1 \to 3a_1$ electron excitation from the carbon center.
Unlike the modeling of valence-excited states, for such a core-excited state we must take into account the orbital relaxation induced by the core vacancy. 
We do this by calculating the DFT and HF molecular orbitals in the presence of a fractional positive charge equally distributed over the two 1$a_1$ spin molecular orbitals, to preserve the symmetry between the two spin channels.
We note that, even at the HF level, the use of fractional occupation in the calculation of the orbital basis breaks away from conventional CI. For instance, with such a basis Brillouin's theorem cease to apply and the fractionally-occupied HF ground state does couple with single-excitation configurations. 
In Figure~\ref{CH4_Core}, we benchmark the core-excited level calculated with both the relaxed cation and non-relaxed neutral ground state molecular-orbital bases against Photoabsorption-spectroscopy experiments by Ueda \textit{et al.}~\cite{Ueda:1995} and Kivimaki~\textit{et al.} \cite{Kivimaki:1996}. Our results at the largest AS are also listed in Table~\ref{Table:CH4}.

Figure~\ref{CH4_Core}~a) shows the vertical core-excitation energy calculated with the neutral non-relaxed HF and DFT orbital bases.
For all our computations, we observe a monotonous convergence of the excitation energy, which reaches a plateau for virtuals above 75 eV.
For all AS, the CI/DFT results are systematically better than the CI/HF analogs when compared to the experimental references~\cite{Ueda:1995,Kivimaki:1996}.
With the DFT basis, the inclusion of the doubles negligibly improves the excitation energy by about 0.6~eV while it has a negligible effect with the HF one -- see the right column of Table~\ref{Table:CH4}. Here as well, we attribute this behavior to the small dynamic correlation affecting CH$_4$, even though the higher sensitivity of the CI/DFT results to the excitation level shows that the the DFT molecular-orbital basis provides a more accurate modeling of core electron-correlation effects compared to the HF equivalent. At the largest AS, our results are 11.3 eV overestimated with respect to the references, thus displaying the insufficiency of the non-relaxed molecular-orbital basis in the modeling of core-excited states even for weakly correlated molecules. 

We now switch to Figure~\ref{CH4_Core}~b), where we compare our calculations with the cation relaxed basis to the references aforementioned. The convergence is slower than the non-relaxed counterparts and it is attained for AS close to the complete-active-space limit. The inclusion of the doubles provides at most marginal  improvement to the excitation energies and the calculations on the DFT molecular orbitals feature a higher sensitivity to the excitation level.
The most striking feature is the excellent agreement with the experimental reference for all CIS/RAS-CISD and HF/DFT bases. This highlights the importance of accounting for orbital relaxation, even in the core-excited state of a poorly correlated system such as CH$_4$, as well as the accuracy of our modeling strategy.

\begin{table}[htb] 
    \caption{\label{Table:CH4}
    Vertical excitation energies for (middle column) valence and (right) C 1s core excitations at the largest AS of figures~\ref{CH4_Valence} and~\ref{CH4_Core}, respectively. For the core excitation, the main numbers (resp. between parentheses) correspond to results using the relaxed-orbital (resp. non-relaxed) basis -- see text.
    For comparison, we reproduce numerical results with [a] complex Kohn calculations by Gil \textit{et al.}~\cite{Gil:1994}, the [b] MRCISD, CASCI and CASSCF computations by Zi\'{o}\l{}kowski \textit{et al.}~\cite{Ziolkowski:2012} ), and experiments by [c]  Ueda \textit{et al.}~\cite{Ueda:1995} and [d] Kivimaki \textit{et al.}~\cite{Kivimaki:1996}.}
    \centering
    \begin{tabular}{lcc}
        \hline
                    & \multicolumn{2}{c}{T [eV]}                  \\
                    & $1t_2 \to 3a_1$   & $1a_1 \to 3a_1$          \\
        \hline
        CIS/HF        & 10.353          & 287.939 (298.376)      \\
        CIS/DFT& 10.536          & 287.65 (297.170)       \\
        RAS-CISD/HF   & 10.332          & 287.595 (298.320)      \\
        RAS-CISD/DFT& 10.386          & 287.044 (296.618)      \\
                                                                \hline
        \multirow{4}{*}{Refs.} 
                    & 11.24 [a]          & 287.05 (Expt. [c])     \\
                    & 10.45 (MRCI [b])   & 287.4 (Expt. [d])      \\
                    & 11.769 (CASCI [b]) &                        \\
                    & 9.42 (CASSCF [b])  &                        \\
        \hline
        \hline
    \end{tabular}
\end{table}

\subsection{Valence- and core-excited states of CO$_2$}

We now move to the carbon dioxide molecule CO$_2$, which is more correlated than CH$_4$ due to the presence of double bonds between the carbon and oxygen atoms. We aim to model (i) the lowest-lying valence excited state and (ii) a core-excited state stemming from excitation out of the C edge. We adapt our AS partitioning to the increased complexity of this molecule, as shown in Figure~\ref{Active_space}~b). Specifically, in all calculations we freeze the core $1\sigma_g$ and $1\sigma_u$ molecular orbitals, which are localized on the two O centers and negligibly improve the modeling of the excited states we are targeting.
A second RAS includes the molecular orbitals from $2\sigma_g$ up to $4\sigma_g$ and we allow single and double excitations from there to the LUMO (2$\pi_u$) and LUMO+1 (5$\sigma_g$), to model core-valence correlation. 
The third RAS includes the outer valence $1\pi_u$, $3\sigma_u$ and $1\pi_g$, from which singles, doubles and triples are allowed to the LUMO and LUMO+1 virtuals, to model valence correlation. The fourth RAS contains the virtuals above LUMO+1, and is systematically increased to include all the virtuals up to 80 eV. Only single excitations are allowed from the occupied orbitals to this partition. We employ this scheme to model the valence- and core-excited states of this molecule.

\begin{figure}[htb]
    \centering
    \includegraphics[width=0.75\linewidth]{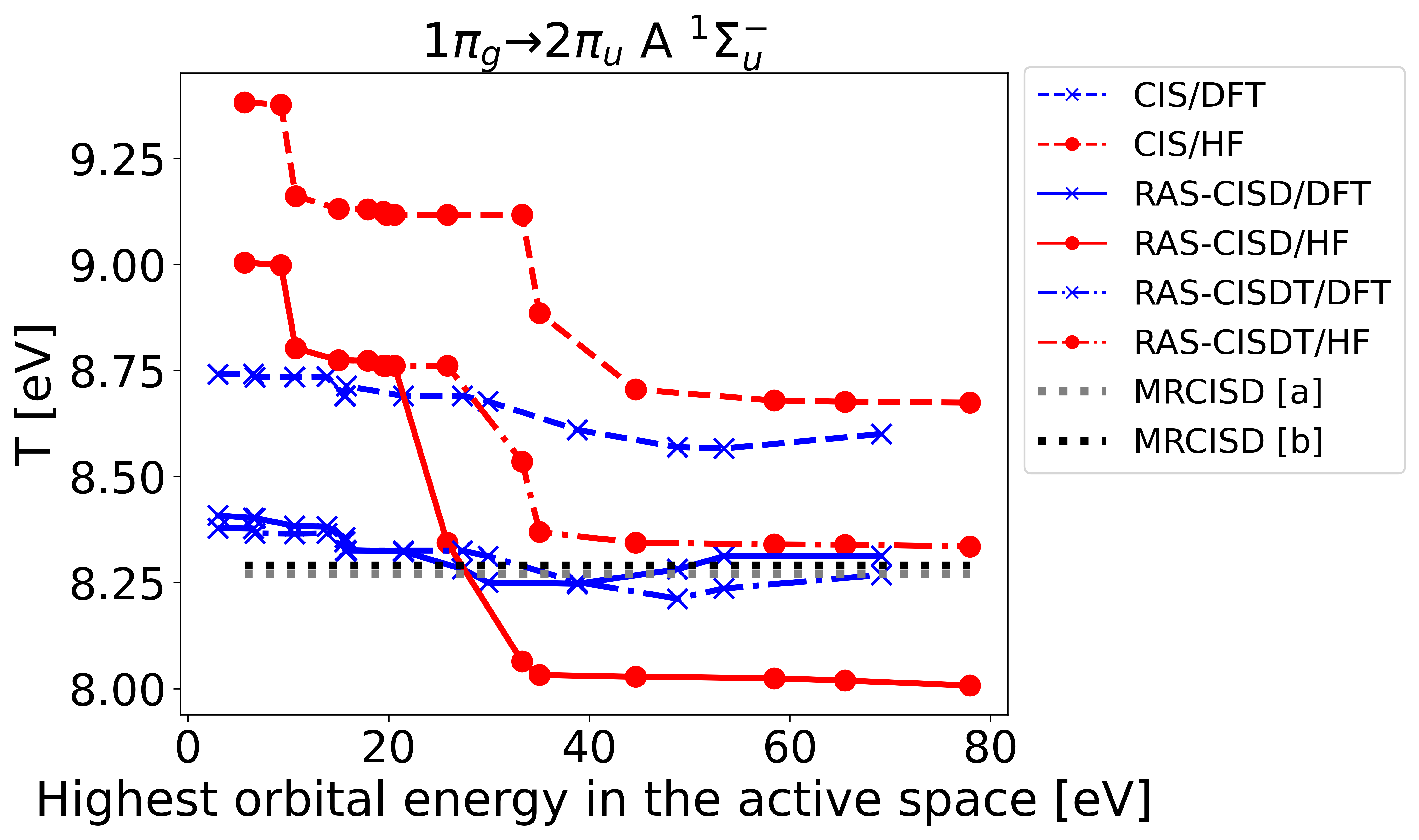}
    \caption{\label{CO2_valence} 
    Convergence pattern for the excitation energies of the A $^1\Sigma_{u}^-$ valence-excited state associated with CO$_2$ in terms of the AS size. We compare our results with [a] the MRCISD calculations on HF molecular orbitals of Winter \textit{et al.}~\cite{Winter:1973} and [b] the MRCISD calculations on multi-configuration self-consistent field (MCSCF) molecular orbitals of Knowles \textit{et al.}~\cite{Knowles:1988}.}
\end{figure}

Figure~\ref{CO2_valence} shows the convergence of the vertical excitation energy $1\pi_g\to2\pi_u$ to the first excited state A$^1\Sigma^{-}_{u}$ in terms of the AS size, and we  summarize the results for the largest AS in Table~\ref{Table: CO2}. 
We benchmark our results against the MRCISD calculations of Winter \textit{et. al.}~\cite{Winter:1973} and Knowles \textit{et. al.}~\cite{Knowles:1988}.
For all calculations, the energy convergence is  attained with virtual orbitals in the AS up to 20~eV for the DFT basis and up to 40~eV for HF.
The inclusion of excitation levels higher than the singles systematically improves the results regardless of the molecular-orbital basis, with the most significant improvement provided by the doubles. In particular, for the DFT basis, the inclusion of the doubles improves the excitation energy by almost 0.4 eV, while the triples changes the excitation energy  by no more than 0.05 eV. For the CI/HF calculations the trend is analogous with respect to the inclusion of the doubles, while the triples significantly improves the excitation energy.
For all AS, we systematically notice a clear improvement in the excitation energy when using the DFT vs HF orbital bases as compared to MRCISD reference calculations~\cite{Winter:1973,Knowles:1988}. In particular, for the largest AS, the RAS-CISD/DFT calculations localize the excitation energy at 8.31 eV, thus only 0.04 eV above the data by Winter \textit{et al.}~\cite{Winter:1973} and 0.02 eV above the analogs by Knowles \textit{et al.}~\cite{Knowles:1988}. It is particularly worth noticing that the RAS-CISD/DFT results are equivalent to the more computationally demanding RAS-CISDT/HF counterparts, which localize the excitation energies at 8.34 eV ({\it i.e.}, 0.07 and 0.05 eV above the MRCI reference calculations). We ascribe the better performance of the CI/DFT over  CI/HF computations and the equivalence of RAS-CISD/DFT and RAS-CISDT/HF results to the improved modeling of dynamic electron correlation at the DFT level of theory.  

\begin{figure}[H]
    \centering
    \includegraphics[width=0.75\linewidth]{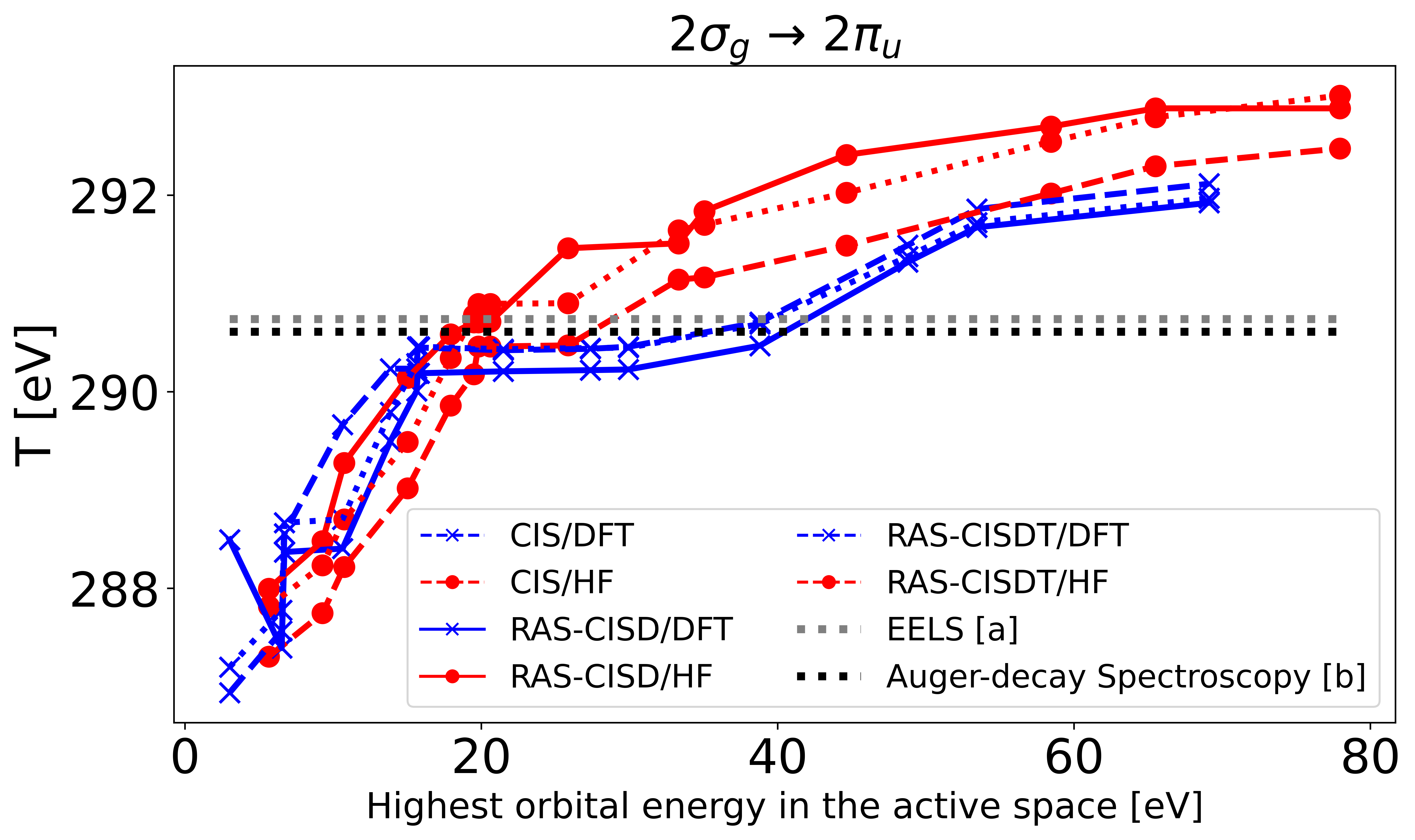}
    \caption{\label{CO2_core-hole} 
    Convergence pattern of the $2\sigma_g\to 2\pi_u$ carbon K edge excitation energy T of CO$_2$ in terms of the active space size. We compare our results with the EELS experiments by Eustatiu \textit{et al.} \cite{Eustatiu:2000} and the Auger-decay spectroscopy data obtained from synchrotron sources by Antonsson \textit{et} \textit{al.} \cite{Antonsson:2015}.}
\end{figure}

We now switch to the modeling of the core-excited state induced by the 2$\sigma_g$ $\to$ 2$\pi_u$ electron transition from the carbon center. 
We follow a similar approach as for CH$_4$ and account for orbital relaxation by calculating the relaxed DFT and HF orbital bases for a CO$_2$ molecule with the same geometry as the neutral but with a +1 charge equally distributed between the two C 1s spin orbitals.
We compare our results with two experimental references for (i) electron-energy-loss spectroscopy (EELS) by Eustatiu \textit{et al.}~\cite{Eustatiu:2000} and (ii) Auger-decay spectroscopy by Antonsson \textit{et al.}~\cite{Antonsson:2015} and show the results in Figure~\ref{CO2_core-hole}. 
We tabulate the results for the largest AS in Table ~\ref{Table: CO2}. 
The CI/DFT calculations attain a first plateau when the AS is extended to virtuals lying at most between 15 and 30 eV. In this region, the excitation energies equal 290.44 eV for CIS/ and RAS-CISDT/DFT and 290.22 for RAS-CISD/DFT and are very close to the experimental references.
When virtual orbital beyond 40~eV are included in the AS, the CI/DFT results deviate markedly from the reference, which we attribute to Rydberg-state contamination from states lying close in energy. 
On the other hand, we do not observe any plateau in the CI/HF calculations. 
Regardless of the molecular-orbital basis, all the calculations stabilize between 65 and 75 eV. At the largest AS, the inclusion of the doubles improves the excitation energies, in particular for the CI/HF calculations, while account for the triples have small or no effects on the accuracy of our results. Nonetheless, the CI/DFT results lie almost 1 eV closer to the experiments than the CI/HF equivalents. These result show that the core-excited states associated with CO\(_2\) feature strong core-orbital relaxation and valence dynamic correlation. The latter is effectively modeled by the interplay between intrinsically correlated DFT molecular orbitals and account for double excitations. Consequently, our RAS-CISD/DFT calculations compete in accuracy with multi-reference methods employing MCSCF orbitals.  

\begin{table}[htb]
    \caption{ \label{Table: CO2}
    Vertical excitation energies for (middle column) valence and (right) C 1s core excitations at the largest AS of figures~\ref{CO2_valence} and~\ref{CO2_core-hole}, respectively. 
    The numbers within parentheses correspond to the excitation energies obtained at the first plateau in the convergence of the CI/DFT calculations -- see text. 
    For comparison, we reproduce numerical results with [a] MRCISD on HF molecular orbitals of Winter \textit{et al.}~\cite{Winter:1973}, [b]  MRCISD calculations on multi-configuration self-consistent field (MCSCF) molecular orbitals of Knowles \textit{et al.}~\cite{Knowles:1988}, and experimental [c]  EELS by Eustatiu \textit{et al.}~\cite{Eustatiu:2000} and [d]  Auger-decay spectroscopy from synchrotron by Antonsson \textit{et al.}~\cite{Antonsson:2015}.}
    \begin{tabular}{lll}
        \hline
        \hline
                    & \multicolumn{2}{c}{T {[}eV{]}}                                                                      \\
                    & $1\pi_g \to 2\pi_u$ & $2\sigma_g \to  2\pi_u$ \\
        \hline
        CIS/HF         & 8.600              & 292.474               \\
        CIS/DFT& 8.674              & 292.115 (290.440)     \\
        RAS-CISD/HF    & 8.007              & 292.882               \\
        RAS-CISD/DFT& 8.313              & 291.921 (290.220)     \\
        RAS-CISDT/HF   & 8.335              & 293.012               \\
        RAS-CISDT/DFT& 8.268              & 291.967 (290.440)     \\
        \hline
        \multirow{2}{*}{Refs.} 
                    & 8.270 (MRCI/HF [a])    & 290.740 (Expt. [c])   \\
                    & 8.290 (MRCI/MCSCF [b]) & 290.610 (Expt. [d])   \\
        \hline
        \hline
    \end{tabular}
\end{table}

\subsection{Valence- and core-excited states of N$_2$}

To conclude our analyses, we look at the nitrogen molecule N$_2$ which, with its triple bond, is the most strongly correlated of the targets we consider in this Paper. 
To model this molecule, we use a scheme  that resembles the one for CH$_4$ with a partition the AS into two restricted active spaces -- see the sketch of Figure \ref{Active_space}~c). The first RAS includes all the occupied orbitals plus the LUMO and LUMO+1 (3$\sigma_u$), to which both singles and doubles are allowed. The second RAS only includes single excitations and is systematically extended to all the virtuals above LUMO+1, up to about 100 eV orbital energy.  
To assess the importance of static correlation in the description of this state, we run both single- and multi-reference CI computations. For the latter, the reference space includes the HF/DFT ground configuration and the twelve configurations stemming from excitations from the HOMO and HOMO-1 (3$\sigma_g$) molecular orbitals to the LUMO molecular orbitals. 

\begin{figure}[H]
    \centering
    \includegraphics[width=0.75\linewidth]{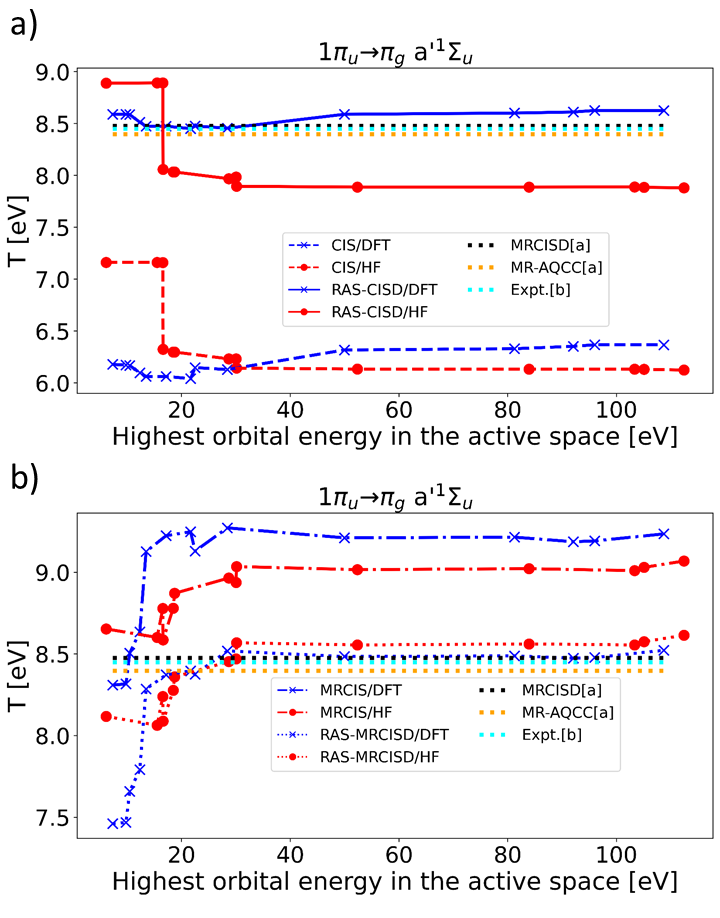}
    \caption{\label{N2_valence}
    Excitation energy $T$ for the $a'{}^{1}\Sigma_{u}$ valence-excited state of N$_2$ computed at the a) single-reference and b) multi-reference CIS and RAS-CISD levels on HF and DFT molecular orbital bases, reported in terms of the binding energy associated with the highest-lying virtual molecular orbital included in the active space. The MRCI and MR-AQCC calculations by M\"{u}ller \textit{et al.}~\cite{Muller:2001} are labelled as [a], whereas the spectroscopic data by Lofthus \textit{et al.}~\cite{Lofthus:1977} are labeled as [b].}
\end{figure}

Figure~\ref{N2_valence} shows the convergence of the vertical excitation energy $1\pi_u\to 1\pi_g$ to the first excited state a'$^1\Sigma_u$ in terms of the highest-lying virtual included in the AS and we tabulate the results at the largest AS in Table~\ref{Table:N2}. 
In particular, Figure~\ref{N2_valence}a) shows the single-reference CI, whereas Figure~\ref{N2_valence}b) reports their multi-reference (MRCI) analogs.
We benchmark our results against  MRCISD and multi-reference average-quadratic coupled cluster (MR-AQCC) calculations on CASSCF molecular orbitals by M\"{u}ller \textit{et al.}~\cite{Muller:2001}, and spectroscopic data of Lofthus \textit{et al.}~\cite{Lofthus:1977}. 
We focus our analysis on three features associated with our calculations: the (i) convergence rate, (ii) excitation level, and (iii) dimension of the reference space.
For all computations convergence is fast and the excitation energies stabilize upon including virtuals beyond 30~eV in the AS. 
At the largest AS, inclusion of the doubles improves the CI/HF and CI/DFT results by 1.8 and 2.3~eV, respectively, as shown in Figure~\ref{N2_valence}a). This shows the higher sensitivity of the DFT basis to the modeling of dynamic electron correlation and is further supported by the better agreement of our CI/DFT excitation energies with the literature data compared to the CI/HF counterparts. 

For both HF and DFT orbital bases, the multi-reference calculations in figure~\ref{N2_valence}~b) show a clear improvement compared to their single-references counterparts in panel~a), with the largest difference obtained for CIS. Specifically, single- and multi-reference CIS calculations differ by almost 3 eV for both the HF and DFT molecular-orbital bases. The case of  RAS-(MR)CISD/DFT is particularly worth of notice, as the single- and multi-reference excitation energies differ by only 0.1 eV -- see the right column of table~\ref{Table:N2}. 
Again, we ascribe the similarity of the single- and multi-reference RAS-CISD/DFT computations to the more accurate modeling of electron-correlation effects within the DFT basis compared to the canonical HF analog. 
In addition, both the single- and multi-reference RAS-(MR)CISD/DFT results agree well with the literature data, with a discrepancy between our calculations and the references being smaller than 0.1~eV. In contrast, accurate performance of the RAS-(MR)CISD/HF calculations requires multi-reference calculations.
Overall, our results are consistent with our observation with CO$_2$ that the electron-correlation effects featured by the valence-excited state are retrieved already by single-reference calculations, provided that a DFT basis is used.

Our last analysis targets the $1\sigma_u\to1\pi_g$ 
core-excited state. Unlike the analogous excited state arising from the lower-lying $1\sigma_g$ molecular orbital, this core-excited state is bright and, therefore, measurements to validate our calculations can be easily found in the literature~\cite{Tronc:1980,King:1977}.
Both the 1$\sigma_g$ and 1$\sigma_u$ molecular orbitals arise from the linear combination of the atoms' 1s atomic orbitals. This feature introduces a higher level of difficulty in modeling as the two atomic centers are equivalent and the molecular orbitals quasi-degenerate. Localization of the hole on either 1$\sigma_g$ or 1$\sigma_u$ leads indeed to transition energies that are in reasonable agreement with the experiments~\cite{Rescigno:1979, Muller:1979, Kryzhevoi:2003}, but at the cost of breaking the symmetry of the total wavefunction~\cite{Rocha:2011}. Our strategy aims to model the K-shell hole without any symmetry breaking, which would decrease the accuracy and physicality of our results. Therefore,  we first optimize the HF and DFT molecular orbitals in the presence of a positive charge equally delocalized over the 1$\sigma_g$ and 1$\sigma_u$  orbitals and ranging from 0 (no electron ejected from the K-shell) to +4 (all electrons ejected from the K-shell). We use these molecular orbitals as bases for MRCIS and RAS-MRCISD calculations.  We benchmark our results against the inner-shell CASSCF (IS-CASSCF) and general multiconfiguration perturbation-theory (IS-GMPCT) calculations of Rocha and De Moura \cite{Rocha:2011}, alongside the EELS experiments by Sodhi and Brion \cite{Sodhi:1984}. 

\begin{figure}[H]
    \centering
    \includegraphics[width=0.75\linewidth]{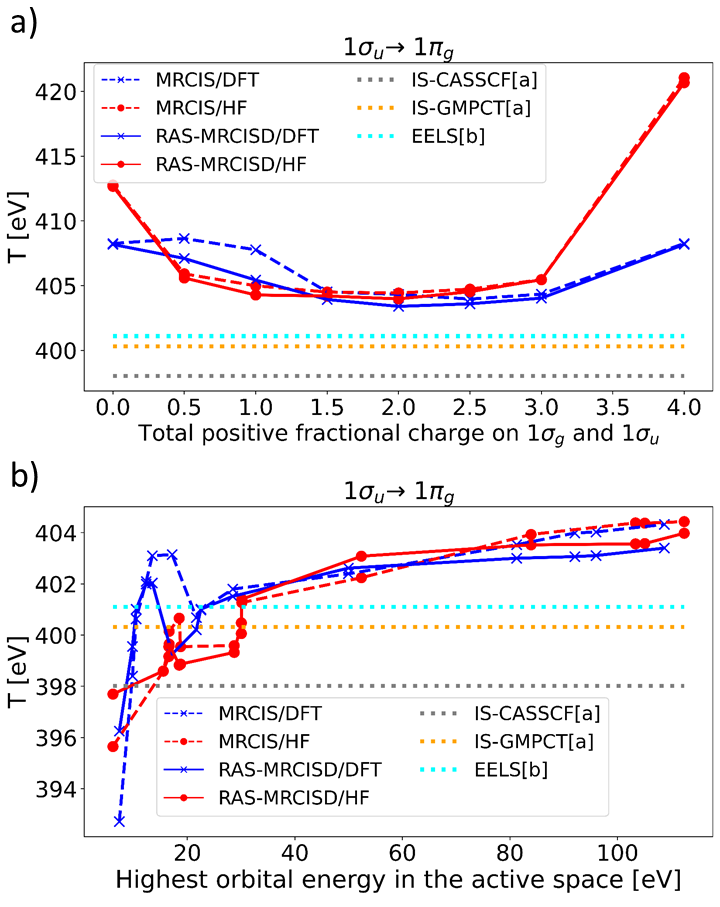}
    \caption{\label{N2_core}
    MRCIS and RAS-MRCISD excitation energies $T$ for the 1$\sigma_u$ $\to$ 1$\pi_g$ core-electron transition of N$_2$ in terms of a) the magnitude of the localized positive charge on the 1$\sigma_g$ and 1$\sigma_u$ molecular orbitals and b) the binding energy associated with the highest-lying virtual included in the active space. Our results are compared with the inner-shell CASSCF (IS-CASSCF) and inner-shell general multi-configuration perturbation-theory (IS-GMCPT) calculations by Rocha and De Moura \cite{Rocha:2011} and the EELS experiments by Sodhi and Brion \cite{Sodhi:1984}. }
\end{figure}

Figure~\ref{N2_core}~a) shows the variation of the core excitation energy $1\sigma_u\to1\pi_g$ at the largest AS as a function of the positive charge on the K-shell.
For all the CI calculations, the excitation energy decreases as the total positive charge on the K-shell increases from +0 to +2.0, while it starts increasing again from +2.5 up to reach its highest value at +4.0.
We ascribe the core-excitation energy lowering to correlation effects on the delocalized basis~\cite{Rocha:2011}.
The variation in excitation energy is more pronounced for HF orbitals than DFT.
The minimum at +2.0 positive charge depicts the physical situation where a hole is formed in the 1$\sigma_u$ molecular orbital upon electron excitation.  Around the +2 charge minimum, DFT basis yields excitation energies that are slightly closer to the references compared to the HF.

Figure~\ref{N2_core}b) shows the convergence of the 1$\sigma_u$ $\to$ 1$\pi_g$ vertical excitation energy as a function of the AS size and for the optimal +2 delocalized charge in the computation of the HF/DFT orbital bases. The convergence pattern is irregular when only including virtual orbitals below 30 eV. Above 80~eV, inclusion of the doubles improves the excitation energy for both the HF and DFT bases.
For all the calculations, once it has stabilized the excitation energy grows with the AS size, as a consequence of the nonphysical mixing of the core-excited state with close-lying Rydberg states that strongly affect the accuracy of our calculations. 
As shown in the right column of Table~\ref{Table:N2}, our RAS-MRCIS/DFT and /HF approaches localize the excitation energy at 404.32 and 404.43 eV, respectively. Inclusion of the doubles decreases these results by slightly more than 1 eV. The improvement related to the flavor of the molecular-orbital basis amounts to no more than 0.6 eV. Thus, even our best results overestimate the EELS measurements by 2.3 eV. Such discrepancies imply that that DFT as well as HF molecular orbitals are not optimally suited for the modeling of  the strong static-correlation effects featured by this state.
Instead, the very good agreement of the IS-GMPCT calculations with the EELS experiments suggests that for excited states with strong multi-reference character our MRCI/DFT approach cannot compete with MRCI calculations on more sophisticated bases optimized for the description of static correlation, such as MCSCF or CASSCF molecular orbitals.

\begin{table}[htb]
    \centering
    \caption{Vertical excitation energies for (middle column) valence and (right) N 1s core excitations at the largest AS of Figures~\ref{N2_valence} and~\ref{N2_core}, respectively. For the valence excitation, the main numbers (resp. between parentheses) correspond to MRCI (CI) calculations -- see text. For comparison, we reproduce numerical results with [a] MRCI and MR-AQCC by M\"{u}ller \textit{et al.}~\cite{Muller:2001}, [b] spectroscopy by Lofthus \textit{et al.}~\cite{Lofthus:1977}, [c] IS-CASSCF and -GMPCT by Rocha and De Moura~\cite{Rocha:2011}, and experimental [d] EELS by Sodhi and Brion~\cite{Sodhi:1984}.}
    \label{Table:N2}
    \begin{tabular}{lll}
        \hline
        \hline
                    & \multicolumn{2}{c}{T {[}eV{]}}               \\
                    & $1\pi_u \to 1\pi_g$ & $1\sigma_u \to 1\pi_g$ \\
        \hline
        MRCIS (CIS)/HF         & 9.069 (6.124)       & 404.430 \\
        MRCIS (CIS)/DFT& 9.235 (6.366)       & 404.321 \\
        RAS-MRCISD (CISD)/HF& 8.614 (7.880)       & 403.977 \\
        RAS-MRCISD (CISD)/DFT& 8.522 (8.625)       & 403.397 \\
        \hline
        \multirow{3}{*}{Refs.} 
                    & 8.476 (MRCI/CASSCF [a]) & 398.02 (IS-CASSCF [c]) \\
                    & 8.396 (MR-AQCC [a])     & 400.32 (IS-GMPCT [c]) \\
                    & 8.449 (Expt. [b])       & 401.10 (Expt. [d]) \\
        \hline
        \hline
    \end{tabular}
\end{table}
%
%
\section{Conclusions} \label{sec:Conclusions}

We investigated a CI approach where the wavefunction is expanded into configurations that are written in a basis of DFT molecular orbitals. We first tested the performance of this CI/DFT approach on the energy convergence of the LiH's ground state.
Our test shows the equivalency of CI/DFT and CI/HF calculations at the full-CI limits, due to the Hermitian nature of the DFT and HF operators. 
Then, we assessed our approach on the valence- and core-excited states of three molecules with increasing modeling complexity, {\it i.e.}, CH$_4$, CO$_2$ and N$_2$. For all these species, account for orbital relaxation is essential to obtain physical core-excitation energies, while the importance of the molecular-orbital basis and excitation levels depends on the related electron-correlation effects. 
For weakly correlated molecules such as CH$_4$, our approach does not provide any significant improvement in the modeling of excited states compared to standard CI/HF calculations. On the other hand, the modeling of  excited states in strongly correlated molecules such as CO$_2$ and N$_2$ is improved at the CI/DFT level and requires the inclusion of double excitations within the CI wave function for accurate description of the excitation energies.
In particular, for the low-lying valence-excited states of CO$_2$ and N$_2$, as well as the 2$\sigma_g \to$ 2$\pi_u$ core-excited state of CO$_2$, which all feature strong dynamic correlation but moderate multi-reference character, our approach is more accurate than the standard CI/HF method and competes with state-of-the-art MRCI/MCSCF calculations. 
In contrast, our approach does not ensure quantitatively accurate excitation energies for excited states featuring strong multi-reference character, such as the 1$\sigma_u \to$ 1$\pi_g$ core-excited state of N$_2$. The modeling of these states rather requires CI calculations on molecular orbitals pre-optimized to account for strong static-correlation effects, like MCSCF and CASSCF molecular orbitals. 
 Future improvements of our approach should thus target the use of CASSCF or MCSCF molecular orbitals optimized from a DFT start~\cite{CASSCF-DFT}.
%

\begin{acknowledgement}

The authors thank Luca Argenti and Kenneth Lopata for enlightening discussions.
This material is based upon work supported by the National Science Foundation under Grant No.~PHY-2207656.

\end{acknowledgement}

\begin{suppinfo}

Extended CI/DFT calculations using the PBE0 and B3LYP DFT functionals. Across calculations, we have systematically found that CI/DFT-PBE0 and -B3LYP yield similar results, leading to curves that are essentially on top of each other in plots. For clarity, in this supporting information, in all plots we omit the CI/DFT-B3LYP curves and report their results at the largest active space limit in tables.

\section{CH$_4$}

Figure~\ref{CH4_ground} adds the ground-state energy convergence for the CI/PBE0 calculations to Figure 3 from the main text. As for Figure 3, the energy error is taken as the difference between the smallest electronic energy at the largest AS and the energy for each active space (AS) size. The CI/PBE0 calculations lie between the CI/BLYP and CI/HF analogs. Its convergence trend mimics the equivalent for the CI/BLYP calculations and converge to very similar energies at the largest AS.

\begin{figure}[htbp]
    \centering
    \includegraphics[width=0.5\linewidth]{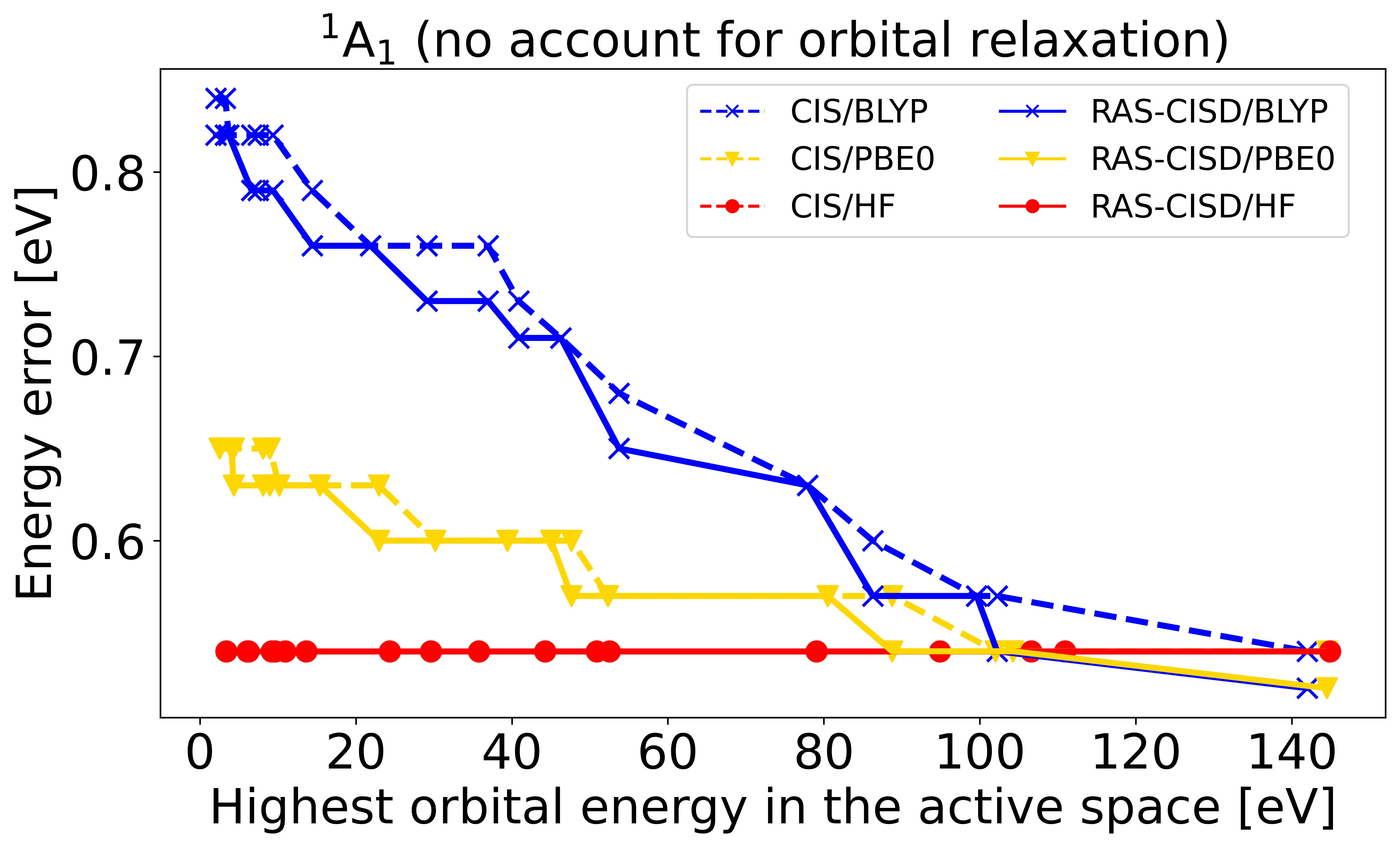}
    \caption{\label{CH4_ground}
    Same as Figure 3 from the main text, with the additional ground-state energy convergence obtained from the CI/PBE0 calculations.
    }
\end{figure}

Figure~\ref{CH4_exc} differs from Figure 4 from the main text for two features: i) the energy error is reported instead of the vertical excitation energy and ii) we added the convergence of the CI/PBE0 calculations to check whether different flavors of the DFT molecular orbitals affect the accuracy of the excitation energy. For this excited state, the energy error is estimated as the difference between our computed excitation energies and the MRCI calculations by Zi\'{o}\l{}kowski \textit{et al.}~\cite{Ziolkowski:2012} -- solid black line. 
We find analogous trends to those shown in Figure~\ref{CH4_ground},  as  the CI/PBE0 results lie between the CI/HF and CI/BLYP equivalent.  At the largest AS, the effect of the functional on the excitation energy is negligible, as the difference between the CI/BLYP and CI/PBE0 results is less than 0.1 eV for the CIS calculations and slightly more than 0.1 eV for the CISD analogs. The excitation energies at the largest AS are tabulated in the middle column of Table~\ref{Table:CH4_SI} with the addition of CI/B3LYP results.

\begin{figure}[htbp]
    \centering
    \includegraphics[width=0.5\linewidth]{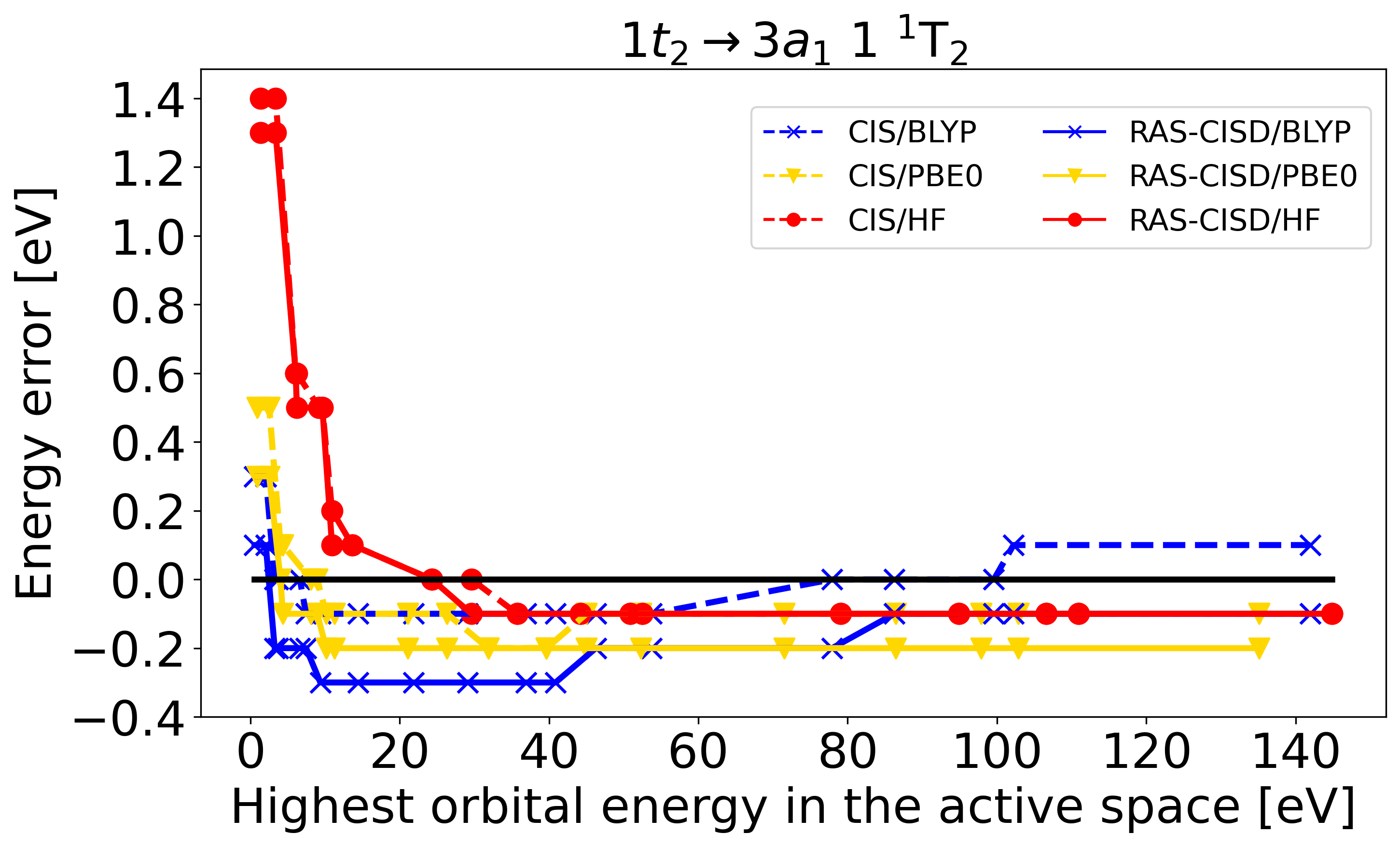}
    \caption{\label{CH4_exc}
   Same as for Figure 4 from the main text, but the convergence pattern is here associated with the energy error for the 1$^1T_2$ state of CH$_4$ in terms of the active-space size. The energy error is taken with respect to the MRCI calculation by  Zi\'{o}\l{}kowski \textit{et al.}~\cite{Ziolkowski:2012} -- the zero black-solid  baseline.
    }
\end{figure}

Figure~\ref{CH4_core_SI} resembles Figure 5 from the main text, but adds the convergence of the CI/PBE0 calculations. Here, the energy error is estimated as the difference between our calculations and the average between the measurements of Ueda \textit{et al.} \cite{Ueda:1995} and Kivimaki \textit{et al.}~\cite{Kivimaki:1996}. For either bases, the CI/PBE0 convergence follows a similar trend to the CI/BLYP analog, but lies closer to the CI/HF calculations. Calculations on the relaxed and non-relaxed bases feature similar differences between the CI/PBE0 and CI/BLYP energy errors at the largest AS. For the CIS level, this difference is less than 0.1 eV regardless of the account for orbital relaxation; for the CISD level, the CI/PBE0 excitation energy obtained with the relaxed basis is higher than the CI/BLYP one by 0.2 eV, while this value increases to 0.5 eV for the non-relaxed basis. 
We list the related excitation energies in the right column of Table~\ref{Table:CH4_SI} , together with the CI/HF and CI/B3LYP analogs. 

\begin{figure}[htbp]
    \centering
    \includegraphics[width=0.5\linewidth]{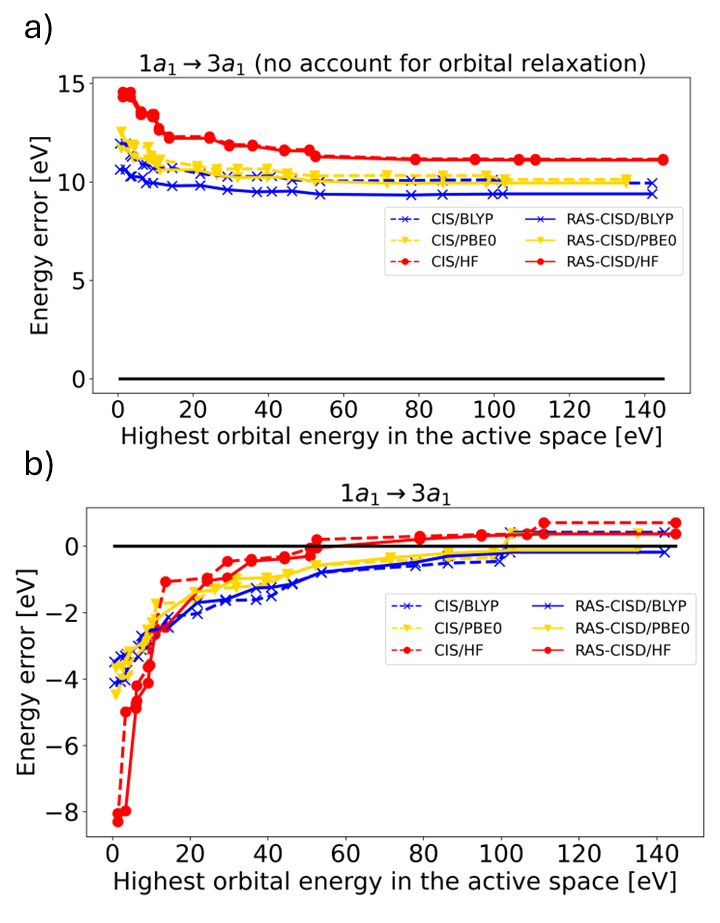}
    \caption{\label{CH4_core_SI}
    Same as for Figure 5 from the main text, but the convergence pattern is here associated with the energy error for the core-excited state arising from the 1$a_1$ $\to$ 3$a_1$ excitation of CH$_4$ in terms of the AS size for the a) non-relaxed and b) relaxed molecular-orbital bases. The energy error is taken with respect to the average of the experimental data by  Ueda \textit{et al.}~\cite{Ueda:1995} and Kivimaki \textit{et al.}~\cite{Kivimaki:1996}(taken as the zero black-solid  baseline in the figure).
    }
\end{figure}

\begin{table}[htbp] 
    \caption{\label{Table:CH4_SI}
    Vertical excitation energies for (middle column) valence and (right) C 1s core excitations at the largest AS of figures~\ref{CH4_exc} and~\ref{CH4_core_SI}, respectively. For the core excitation, the main numbers (resp. between parentheses) correspond to results using the relaxed-orbital (resp. non-relaxed) basis -- see main text.
    For comparison, we reproduce numerical results with [a] complex Kohn calculations by Gil \textit{et al.}~\cite{Gil:1994}, the [b] MRCISD, CASCI and CASSCF computations by Zi\'{o}\l{}kowski \textit{et al.}~\cite{Ziolkowski:2012} ), and experiments by [c]  Ueda \textit{et al.}~\cite{Ueda:1995} and [d] Kivimaki \textit{et al.}~\cite{Kivimaki:1996}.}
    \centering
    \begin{tabular}{lcc}
        \hline
                    & \multicolumn{2}{c}{T [eV]}                  \\
                    & $1t_2 \to 3a_1$   & $1a_1 \to 3a_1$          \\
        \hline
        CIS/HF        & 10.353          & 287.939 (298.376)      \\
 CIS/PBE0& 10.349&287.59 (297.35)\\
 CIS/B3LYP& 10.404&286.87 (297.51)\\
        CIS/BLYP& 10.536          & 287.65 (297.170)       \\
        RAS-CISD/HF   & 10.332          & 287.595 (298.320)      \\
 RAS-CISD/PBE0& 10.264&287.10 (297.17)\\
 RAS-CISD/B3LYP& 10.258&287.14 (296.44)\\
        RAS-CISD/BLYP& 10.386          & 287.044 (296.618)      \\
                                                                \hline
        \multirow{4}{*}{Refs.} 
                    & 11.24 [a]          & 287.05 (Expt. [c])     \\
                    & 10.45 (MRCI [b])   & 287.4 (Expt. [d])      \\
                    & 11.769 (CASCI [b]) &                        \\
                    & 9.42 (CASSCF [b])  &                        \\
        \hline
        \hline
    \end{tabular}
\end{table}

\section{CO$_2$}

Figure~\ref{CO2_ground} shows the energy error for the ground-state energy convergence of CO$_2$ in terms of the highest virtual orbital within the AS. As in Figure~\ref{CH4_ground}, we plotted our CI calculations on HF, BLYP and PBE0 molecular orbitals and assess the effect of the functional on the ground-state energy convergence. In analogy with the ground state of CH$_4$, the flavor of the DFT molecular-orbital basis has no remarkable influence on the energy convergence at the largest AS. 
at the CIS level the two CI/DFT calculations differ by 0.3 eV.  Similarly, at the CISD and CISDT levels, the choice of the functional for the molecular orbitals makes no difference in the ground-state energy.

\begin{figure}[htbp]
    \centering
    \includegraphics[width=0.5\linewidth]{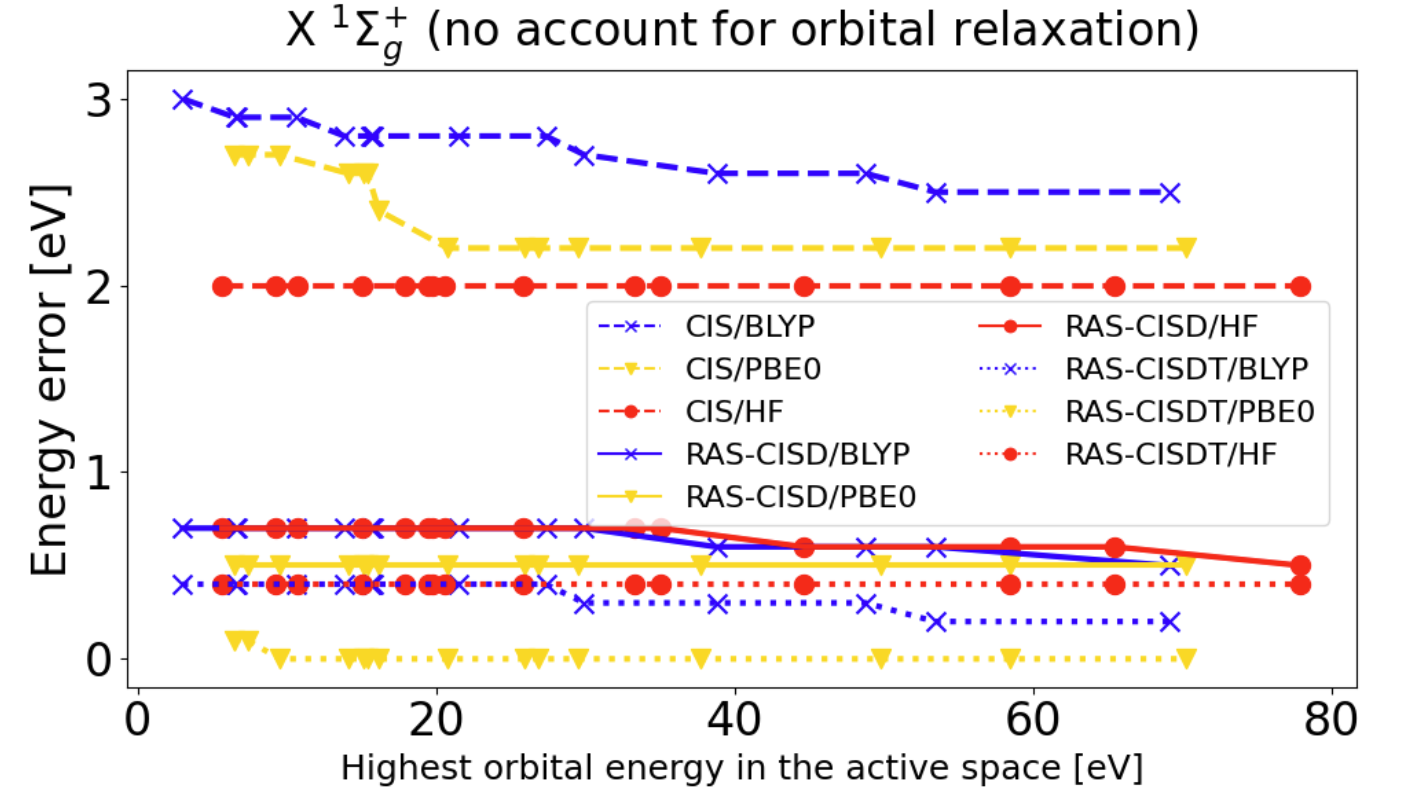}
    \caption{\label{CO2_ground}
    Convergence pattern associated with the ground state of CO$_2$ in terms of the active-space size for the a) non-relaxed and b) relaxed molecular-orbital bases.
    }
\end{figure}

Figure~\ref{CO2_exc} displays the energy error as the difference between the vertical excitation energies shown in Figure 6 from the main text and the average of the MRCI calculations by Winter \textit{et al.}~\cite{Winter:1973} and Knowles \textit{et al.}~\cite{Knowles:1988} -- black solid baseline. In addition, we estimated the energy error associated with the CI/PBE0 calculations.  The trends for these states mirror the analogs for the excited states of CH$_4$. At the largest AS, the CIS/BLYP and CIS/PBE0 excitation energies differ by less than 0.1 eV.  This difference rises to 0.1 eV upon including the doubles into the CI calculations, whereas inclusion of the triples decreases the functional-related energy spread back to less than 0.1 eV.  The related excitation energies at the convergence limit are listed in the middle column of~\ref{Table: CO2_SI}.

\begin{figure}[htbp]
    \centering
    \includegraphics[width=0.5\linewidth]{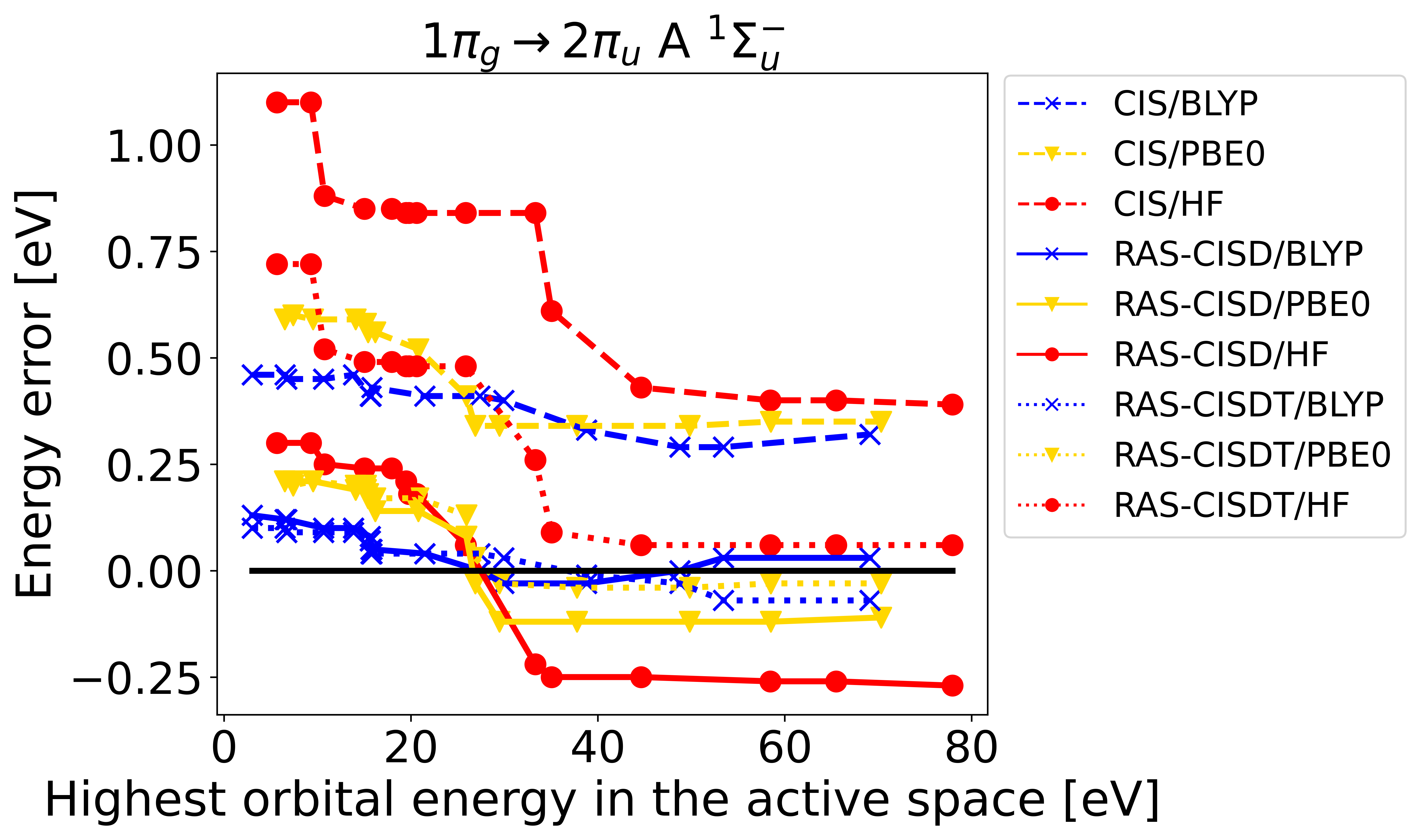}
    \caption{\label{CO2_exc}
  Same as for Figure 6 from the main text. Here, the convergence pattern is associated with the energy error for the A$^1\Sigma^-_u$ state of CO$_2$ in terms of the active-space size. The energy error is taken with respect to the average of the MRCI calculations by  Winter \textit{et al.}~\cite{Winter:1973} and Knowles \textit{et al.}~\cite{Knowles:1988} -- the zero black-solid  baseline.
    }
\end{figure}

Figure~\ref{CO2_core_SI} plots the energy error associated with the vertical excitation energies shown in Figure 7 from the main text, plus the analog for the CI/PBE0 calculations. The energy error is evaluated as the difference between our results and he average of the experiments by Eustatiu \textit{et al.}~\cite{Eustatiu:2000} and Antonsson \textit{et al.}~\cite{Antonsson:2015} (see main text for details on the experimental techniques). The excitation energies obtained at the largest AS are also listed in the right column of Table~\ref{Table: CO2_SI}.
For each size of the AS, the CI/PBE0 calculations lie between the CI/BLYP and CI/HF analogs, in analogy with the convergence of the valence-excited state. Between 15 and 30 eV, the CI/B3LYP calculations exhibit a plateau analogous to the one reported for the CI/BLYP analogs in the main text, while this feature is  absent in the convergence of the CI/PBE0 excitation energies.  We speculate that this behavior is due to the different amount of HF exchange functional within the hybrid DFT functionals which, in turn, determines the mixing with close-lying Rydberg states in this AS region. 
Our interpretation is supported by the similar trend shown by the CI/HF computations. The effect of the functional on the excitation energy is larger for this core-excited state compared to the analogous feature for the valence-excited state, but still small. For the CIS and CISDT levels, the energy difference due to the BLYP and PBE0 functionals amount to 0.1 and 0.2~eV, respectively. This value increases to about 0.6 eV for the CISD level -- see Table~\ref{Table: CO2_SI}.

\begin{figure}[htbp]
    \centering
    \includegraphics[width=0.5\linewidth]{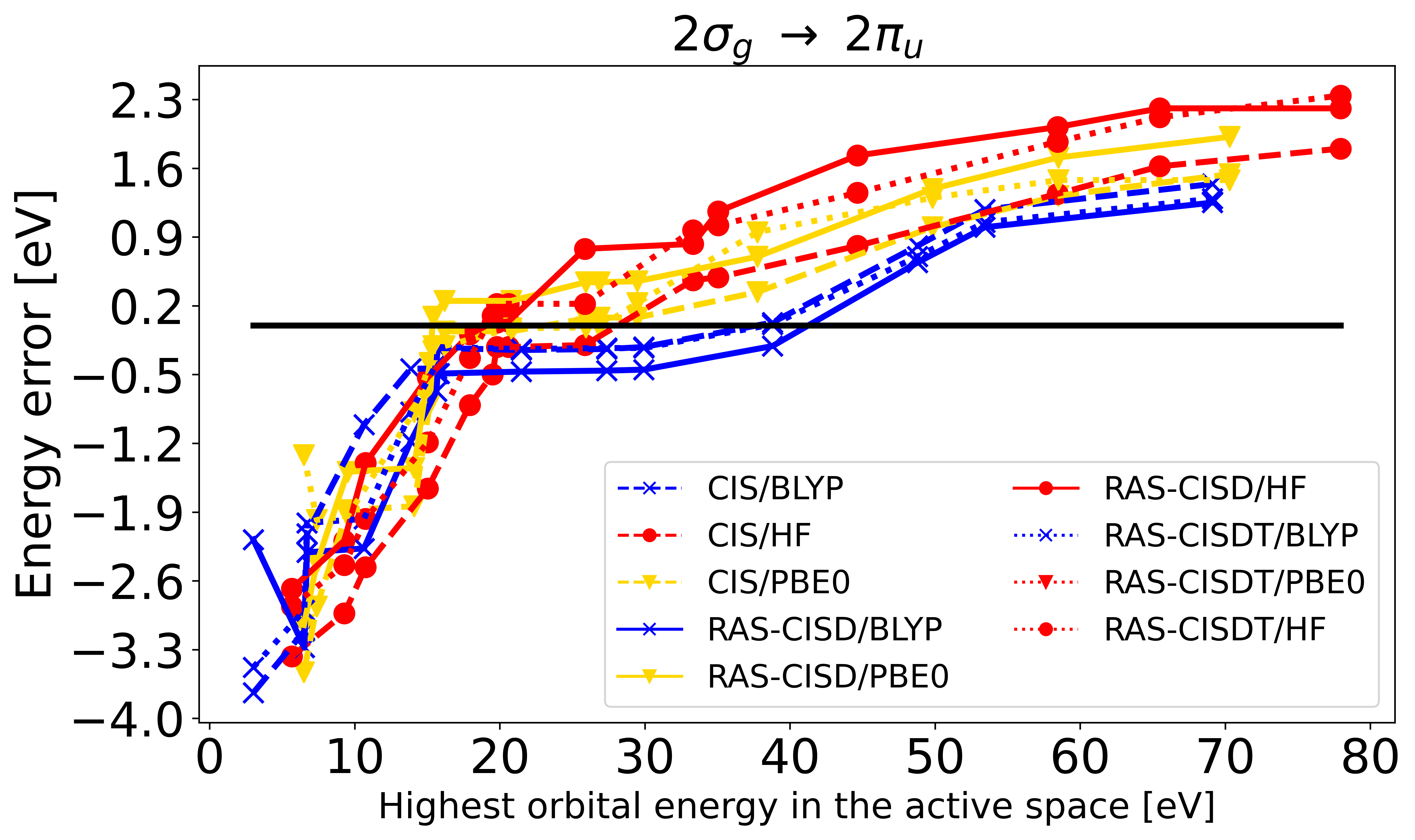}
    \caption{\label{CO2_core_SI}
    Same as for Figure 7 from the main text. Here, the convergence pattern is associated with the energy error for the core-excited state arising from the 2$\sigma_g$ $\to$ 2$\pi_u$ excitation of CO$_2$ in terms of the active-space size. The energy error is taken with respect to the average of the experimental data by  Eustatiu \textit{et al.}~\cite{Eustatiu:2000} and Antonsson \textit{et al.}~\cite{Antonsson:2015}(taken as the zero black-solid  baseline in the figure).
    }
\end{figure}

\begin{table}[htbp]
    \caption{ \label{Table: CO2_SI}
    Vertical excitation energies for (middle row) valence and (right) C 1s core excitations at the largest AS of figures~\ref{CO2_exc} and~S\ref{CO2_core_SI}, respectively. 
    The numbers within parentheses correspond to the excitation energies obtained at the first plateau in the convergence of the CI/DFT calculations -- see text. 
    For comparison, we reproduce numerical results with [a] MRCISD on HF molecular orbitals of Winter \textit{et al.}~\cite{Winter:1973}, [b]  MRCISD calculations on multi-configuration self-consistent field (MCSCF) molecular orbitals of Knowles \textit{et al.}~\cite{Knowles:1988}, and experimental [c]  electron-energy-loss spectroscopic  (EELS) by Eustatiu \textit{et al.}~\cite{Eustatiu:2000} and [d]  Auger-decay spectroscopy from synchrotron by Antonsson \textit{et al.}~\cite{Antonsson:2015}.}
    \begin{tabular}{lll}
        \hline
        \hline
                    & \multicolumn{2}{c}{T {[}eV{]}}                                                                      \\
                    & $1\pi_g \to 2\pi_u$ & $2\sigma_g \to  2\pi_u$ \\
        \hline
        CIS/HF         & 8.600              & 292.474               \\
 CIS/PBE0& 8.632&292.215\\
 CIS/B3LYP& 8.631&292.183 (290.488)\\
        CIS/BLYP& 8.674              & 292.115 (290.440)     \\
        RAS-CISD/HF    & 8.007              & 292.882               \\
 RAS-CISD/PBE0& 8.170&292.598 (290.346)\\
 RAS-CISD/B3LYP& 8.435&292.130\\
        RAS-CISD/BLYP& 8.313              & 291.921 (290.220)     \\
        RAS-CISDT/HF   & 8.335              & 293.012 (290.819)\\
 RAS-CISDT/PBE0& 8.254&292.159\\
 RAS-CISDT/B3LYP& 8.246&292.575\\
        RAS-CISDT/BLYP& 8.268              & 291.967 (290.440)     \\
        \hline
        \multirow{2}{*}{Refs.} 
                    & 8.270 (MRCI/HF [a])    & 290.740 (Expt. [c])   \\
                    & 8.290 (MRCI/MCSCF [b]) & 290.610 (Expt. [d])   \\
        \hline
        \hline
    \end{tabular}
\end{table}

\end{suppinfo}


\providecommand{\latin}[1]{#1}
\makeatletter
\providecommand{\doi}
  {\begingroup\let\do\@makeother\dospecials
  \catcode`\{=1 \catcode`\}=2 \doi@aux}
\providecommand{\doi@aux}[1]{\endgroup\texttt{#1}}
\makeatother
\providecommand*\mcitethebibliography{\thebibliography}
\csname @ifundefined\endcsname{endmcitethebibliography}  {\let\endmcitethebibliography\endthebibliography}{}

\end{document}